\documentclass[%
reprint,
superscriptaddress,
nobibnotes,
amsmath,amssymb,
aps,
pre,
]{revtex4-2}

\usepackage{CJK}%
\usepackage{graphicx}%
\usepackage{dcolumn}%
\usepackage{bm}%
\usepackage{hyperref}
\hypersetup{breaklinks=true,   %
            colorlinks=true,
            linkcolor=blue,
            citecolor=blue,
            urlcolor=blue}%

\usepackage{soul}%
\usepackage{mathtools}%

\begin{document}

\begin{CJK*}{UTF8}{gbsn}

\title{Controlling tissue size by active fracture}%

\author{Wei Wang (汪巍)}
\affiliation{%
Department of Physics and Astronomy, Johns Hopkins University, Baltimore, Maryland 21218, USA\\
}%

\author{Brian A. Camley}
\affiliation{%
Department of Physics and Astronomy, Johns Hopkins University, Baltimore, Maryland 21218, USA\\
}%
\affiliation{%
Department of Biophysics, Johns Hopkins University, Baltimore, Maryland 21218, USA\\
}

\begin{abstract}
Groups of cells, including clusters of cancerous cells, multicellular organisms, and developing organs, may both grow and break apart. What physical factors control these fractures? In these processes, what sets the eventual size of clusters?
We {first develop a one-dimensional framework} for understanding cell clusters that can fragment due to cell motility using an active particle model.
We compute analytically how the break rate of cell-cell junctions depends on cell speed, cell persistence, and cell-cell junction properties.
Next, we find the cluster size distributions, which differ depending on whether all cells can divide or only the cells on the edge of the cluster divide.
Cluster size distributions depend solely on the ratio of the break rate to the growth rate---allowing us to predict how cluster size and variability depend on cell motility and cell-cell mechanics.
Our results suggest that organisms can achieve better size control when cell division is restricted to the cluster boundaries or when fracture can be localized to the cluster center.
{Additionally, we derive a universal survival probability for an intact cluster $S(t)=\mathrm{e}^{-k_d t}$ at steady state if all cells can divide, which is \emph{independent} of the rupture kinetics and depends solely on the cell division rate $k_d$.}
{Finally, we further corroborate the one-dimensional analytics with two-dimensional simulations, finding quantitative agreement with some---but not all---elements of the theory across a wide range of cell motility.}
Our results link the general physics problem of a collective active escape over a barrier to size control, providing a quantitative measure of how motility can regulate organ or organism size.
\end{abstract}

\maketitle
\end{CJK*}

\section{Introduction}\label{sec:intro}

How does an organ or an organism control its size? Size is thought to be tightly regulated by feedbacks controlling cell division~\cite{lecuit2007orchestrating}. However, two recent experiments suggest a different possibility---that the size of a group of cells can arise from a competition between growth, which tends to make the group larger, and random cell motility, which can make the group fracture into multiple pieces, reducing group size. In the metazoan \textit{Trichoplax adhaerens}, asexual reproduction by fission is driven by motility-induced  fractures~\cite{prakash2021motility}. %
Similarly, germline cysts in mice are formed by a combination of cell division and fracture of intercellular bridges by random cell motility~\cite{levy2024tug}, {as shown in Fig.~\ref{fig:cysts}}. {These mechanisms, which depend on material failure, are reminiscent of how cancerous cells break from an invading front~\cite{mukherjee2021cluster,law2023cytokinesis,wang2025confinement}---which seems very different than a well-regulated organism attempting to keep a fixed size}.
{Analogous fracture-limited size control also occurs in multicellular snowflake yeast clusters, though driven by crowding-induced mechanical stress, not motility~\cite{jacobeen2018cellular}. 
We want to know: can motility-driven fracture alone reliably regulate the size of a group of cells?} Both \textit{T. adhaerens} and germline cysts show significant variability in size~\cite{prakash2021motility,levy2024tug, ikami2023branched, lei2013mouse}. What physical factors control the group size and its variability?

We argue that the size of cell clusters controlled by fracture is set by competition between the break rate of cell-cell junctions $k_b$ and the cell division rate $k_d$~\cite{nanda2024dynamic}.
We model break rate $k_b$ from a mechanical perspective, using a simple one-dimensional model of cells as active particles connected by springs (Sec.~\ref{sec:mechanics}). The break rate depends on typical cell speed, cell-cell junction strength, and the cell's persistence time. We then develop models of cluster growth and fracture. We derive the exact steady-state cluster size distribution, finding that cluster sizes depend solely on the ratio $k_b/k_d$, allowing us to link cluster size to cell adhesion and motility (Section~\ref{sec:growth_model}). The quality of cluster size control can be improved if only cells on the edge of the cluster divide or if only cell-cell junctions near the cluster middle can fracture.
{{Section~\ref{sec:combine} shows how the model parameters are constrained by available experimental data, and Sec.~\ref{sec:survival} demonstrates that the survival probability of an intact cluster where all cells divide follows the universal decay $S(t)=\mathrm{e}^{-k_d t}$, governed solely by the cell division rate. Finally, we test the robustness of the theory by comparing with two-dimensional active-particle simulations in Sec.~\ref{sec:2d}.}

\section{Results for 1D chain}\label{sec:1d}
\subsection{Mechanics}\label{sec:mechanics}
We initially model our cell cluster as a one-dimensional chain of cells. This geometry is natural for germline cysts {which are composed of potentially-branched chains of cells that can be fragmented by a single junction break (Fig.~\ref{fig:cysts}), and linear geometries of strings of cells are} often found in cells confined in extracellular matrix {or experiments performed in microchannels or micropatterns} ~\cite{law2023cytokinesis, wang2025confinement, desai2013contact, jain2020role}. One dimension may be appropriate for \textit{T. adhaerens} when it takes on an elongated string-like shape~\cite{prakash2021motility}, though a higher-dimensional model is necessary for a full study of \textit{T. adhaerens}.
 We treat our chain of cells as self-propelled active particles with positions $\{x_n\}$ connected by springs [Fig.~\ref{fig:mechanics}(a)], assuming an overdamped environment:
\begin{equation}\label{eq:EOM}
    \dot{x}_n = -\mu \nabla_n\Phi +v_n(t),
\end{equation}
where $\mu$ is the particle mobility and $\Phi=\frac{1}{2}k\sum_{\langle i,j\rangle} \left(|x_i-x_j|-\ell_0\right)^2$ is the cell-cell interaction energy. Here, $k$ is the spring constant, $\ell_0$ is the natural length of the springs, and $\langle i,j\rangle$ denotes the summation over nearest neighbors.
The active velocities $v_n$---the velocities the cell would have in the absence of cell-cell interactions---are  Ornstein-Uhlenbeck (OU) processes~\cite{dunn1987unified}: 
\begin{equation}
    \tau \dot{v}_n = -v_n+\sqrt{2D}\xi_n(t),
\end{equation}
where $\xi_n(t)$ are independent, zero-mean, unit-variance Gaussian white noises. $\tau$ is the persistence time of the cell, while $D$ controls the typical cell speeds---$v_n$ have mean zero and  correlations $\langle v_n(t)v_{n'}(t') \rangle= \delta_{nn'}(D/\tau)\mathrm{e}^{-|t-t'|/\tau}$ if $t,t' \gg \tau$. %
In the limit $\tau\to 0$, the active velocities $v_n$ reduce to Gaussian white noises with correlations $\langle v_n(t)v_{n'}(t') \rangle=\delta_{nn'}2D\delta(t-t')$---in this limit the system is in thermal equilibrium, but will be out of equilibrium for a finite $\tau$.

\begin{figure}
\includegraphics[width=0.4\textwidth]{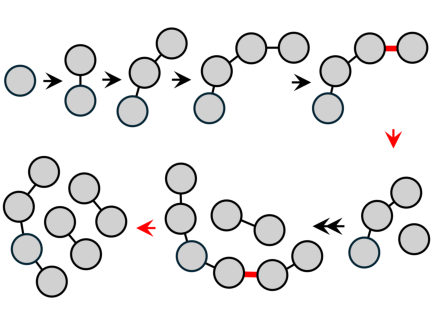}
\caption{{Schematic illustration of cell division and fracture. Red lines denote intercellular bridges that undergo fracture.
}}
\label{fig:cysts}
\end{figure}

What is the mean first time to rupture for a given link, i.e., the time required for the spring to be stretched to a specified threshold?
Our initial intuition was that the rupture rate within a cluster would be the same as that of a two-particle link, as in thermal equilibrium. To explore this, we first compute the variance of the stretched length $\delta\ell=\ell-\ell_0$ (where $\ell$ is the distance between the particles) in the two-particle case.
Extending the approach of Ref.~\cite{woillez2020active}, we can map this problem to an inertial Brownian particle in a harmonic potential $U(\delta\ell)=\mu k\delta\ell^2$ experiencing a friction $\eta=1+2\mu k\tau$, and an effective temperature $k_BT_\textrm{eff}=2D/\eta$ (see Appendix~\ref{app:two_particle}). The distribution of $\delta\ell$ in the steady state is  Boltzmann-like  $\sim\exp(-U/k_B T_\textrm{eff})$, i.e., a Gaussian distribution with zero mean and variance 
\begin{equation}
  \langle\delta\ell^2\rangle_2=\frac{k_BT_\mathrm{eff}}{U''}=\frac{D}{\mu k (1+2\mu k\tau)}.
\end{equation}
The subscript indicates this is the two-particle result.
When $\tau=0$, $\langle\delta\ell^2\rangle_2$ approaches the thermal equilibrium solution $\langle\delta\ell^2\rangle_\mathrm{th}=D/\mu k$. %
If the spring breaks when stretched beyond a critical length {$\ell_0+\delta\ell_b$}, the mean escape time can be estimated using standard Kramers' theory. For small effective temperatures, the time for the pair to break is given by 
\begin{equation}
\tau_\textrm{esc}^\textrm{pair}=\tau_0\exp\left(\frac{\Delta U}{k_B T_\mathrm{eff}}\right)=\tau_0 \exp\left[\frac{\mu k\delta\ell_b^2(1+2\mu k\tau)}{2D}\right],
\end{equation}
 where $\tau_0$ is a subexponential correction~\cite{woillez2020active, gardiner2009stochastic, hanggi1990reaction, haunggi1994colored, wexler2020dynamics}.

{For a long chain of $N\gg 1 $ cells, we assume that $x_n(t)=n\ell_0+\delta x_n(t)$, where $\delta x_n$ represents the deviation of $x_n$ around the lattice point $n\ell_0$. We also assume periodic boundary conditions, $\delta x_{n+N} = \delta x_{n}$---we find this does not influence our central results in the limit $N \to \infty$ and makes calculation much simpler. We use a standard discrete Fourier transform:
\begin{equation}
    \tilde{x}_q=\frac{1}{N}\sum_{n=0}^{N-1}\delta x_n\mathrm{e}^{-iqn},
\end{equation}
where $q=2\pi m/N$ and the index $m=0,1,\cdots,N-1$. The inverse transform is thus given by $\delta x_n=\sum_q\tilde{x}_q\mathrm{e}^{iqn}$.
By substituting the inverse transforms into Eq.~\eqref{eq:EOM} and matching terms for each mode $q$, we can then obtain (see Appendix~\ref{app:chain} for more details):
\begin{equation}\label{eq:EOM_q}
    \dot{\tilde{x}}_q = -\mu k_q\tilde{x}_q+ \tilde{v}_q,
\end{equation}
where $k_q=2k(1-\cos q)$ and $\tilde{v}_q=(1/N)\sum_{n=0}^{N-1}v_n\mathrm{e}^{-iqn}$ is the Fourier amplitude of $v_n$, with mean $\langle \tilde{v}_q\rangle = 0$, and time correlations:
\begin{equation}
    \langle \tilde{v}^{\phantom{\ast}}_q(t)\tilde{v}^\ast_{q'}(t')\rangle = \delta_{qq'}\frac{D}{N\tau}\mathrm{e}^{-|t-t'|/\tau}.
\end{equation}
By solving Eq.~\eqref{eq:EOM_q}, we find that the zeroth mode (center of mass) evolves as $\langle| \tilde{x}_0(t)|^2\rangle = 2D t/N$, while for the nonzero modes:
\begin{equation}
    \langle \tilde{x}^{\phantom{\ast}}_q(t)\tilde{x}^\ast_{q'}(t)\rangle =\delta_{qq'}\frac{D}{N\mu k_q(1+\mu k_q\tau)}.
\end{equation}
We can then calculate the stretched lengths of the springs by $\delta\ell_n = x_{n+1} - x_n-\ell_0=\delta x_{n+1}-\delta x_n=\sum_{q\neq 0}\tilde{x}_q\mathrm{e}^{iqn}(\mathrm{e}^{iq}-1)$, where the mean is simply $\langle\delta\ell_n\rangle = 0$, and the variance is
\begin{eqnarray}
    \langle \delta\ell_n^2\rangle
    &=&\sum_{q\neq 0 }\sum_{q'\neq 0}\langle \tilde{x}^{\phantom{\ast}}_q\tilde{x}^\ast_{q'}\rangle \mathrm{e}^{i(q-q')n}(\mathrm{e}^{iq}-1)(\mathrm{e}^{-iq'}-1)\nonumber\\
    &=&\frac{D}{N\mu k}\sum_{q\neq 0}\frac{1}{1+2\mu k\tau (1-\cos q)}.\nonumber
\end{eqnarray}
Replacing the sum by an integral, we obtain
}
\begin{equation}
    \langle \delta\ell_n^2\rangle = \frac{D}{\mu k\sqrt{1+4\mu k\tau}} ,
\end{equation}
which is different from the two-particle result $\langle\delta\ell^2\rangle_2$ at large $\mu k \tau$ [Fig.~\ref{fig:mechanics}(b)].
\begin{figure}
\includegraphics[width=0.48\textwidth]{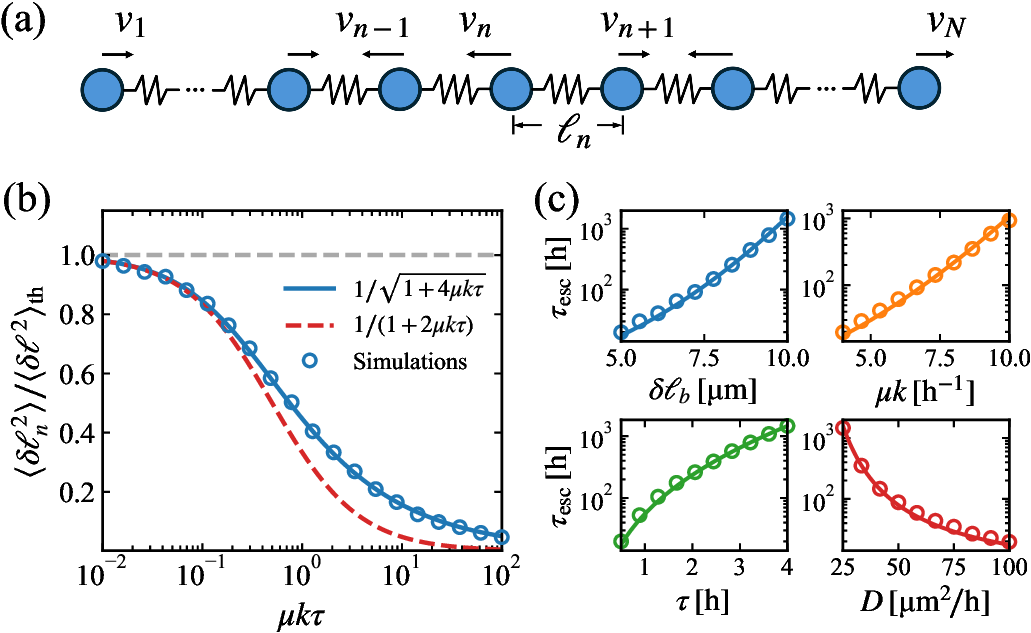}
\caption{
(a) Illustration of the active chain model. Cells are connected by springs; each cell has active velocity $v_n$.
(b) Variance of the spring stretch $\langle\delta\ell_n^2\rangle$ as a function of $\mu k\tau$. Gray dashed line is the thermal variance $\langle\delta\ell^2\rangle_\mathrm{th}=D/\mu k$. We change $\mu k\tau$ by varying $\tau$ while fixing $\mu k$ in the simulations.
Red dashed line is the two-particle result $\langle\delta\ell^2\rangle_2$. %
(c) Mean escape time $\tau_\mathrm{esc}=1/k_b$.
Empty circles are simulation results with $N=1000$ cells; solid lines are theory, Eq.~\eqref{eq:break_rate}.}
\label{fig:mechanics}
\end{figure}
Notably, $\langle\delta\ell_n^2\rangle=\langle\delta\ell^2\rangle_\mathrm{th}$ when $\tau\to0$, but $\langle\delta\ell_n^2\rangle$ also converges to the two-particle result $\langle\delta\ell^2\rangle_2$ when $0<\mu k\tau\ll 1$, as shown in Fig.~\ref{fig:mechanics}(b). This implies that when $\mu k\tau$ is small but finite, the system resides in an \emph{effective equilibrium} regime~\cite{fodor2016far}, and such a convergence can be understood through a modified equipartition theorem (see Appendix~\ref{app:equipartition}).
In the steady state, the distribution of $\delta\ell_n$ follows a Gaussian form, $P_s(\delta\ell_n)\sim\exp(-\delta\ell_n^2/2\langle\delta\ell_n^2\rangle)$.
{The escape rate will be proportional to the probability density at the breaking length $P_s(\delta \ell_b)$~\cite{woillez2020active}, yielding%
\begin{equation}\label{eq:break_rate}
    k_b = \tau_0^{-1}\exp\left(-\frac{\mu k\delta\ell_b^2\sqrt{1+4\mu k\tau}}{2D}\right),
\end{equation}
where we have found the subexponential correction $\tau_0=\pi\eta/\mu k$ leads to a good fit to the escape time from our simulation of active cell motion [see Fig.~\ref{fig:mechanics}(c)].}
Time to rupture increases with a higher break threshold $\delta\ell_b$, stiffer springs, or greater cell persistence, while it decreases if cells are faster (larger $D$).
Unlike the two-particle problem, where the mean escape time grows exponentially with $k$ in equilibrium and $k^2$ in the $\mu k \tau \gg 1$ nonequilibrium limit, it grows here exponentially in $k^{3/2}$ for large $k$.

\subsection{Growth models}\label{sec:growth_model}

We have described an active chain in which links can break at a rate $k_b$, while cells can also divide, leading to an increase in the chain length.
What controls the cluster size, and how broad is its distribution?
As in germline cysts, where all cells can divide~\cite{levy2024tug}, we first assume that each cell divides independently with the same division rate $k_d$.
To analyze the distribution of the chain length, we randomly pick one of the daughter chains to track once fragmentation occurs. The selected daughter chain then continues to grow and rupture in the same manner as the original chain, as illustrated in Fig.~\ref{fig:schematics}(a).

Assuming $p_n(t)$ gives the probability that the length of the tracked chain at time $t$ is $n$, the master equation governing such a process is, generalizing {models of  polymerization and fragmentation}~\cite{krapivsky2010kinetic},
\begin{equation}\label{eq:master_all}
    \frac{\mathrm{d} p_n}{\mathrm{d} t}=\sum_{m>n} k_b p_{m} -p_n\sum_{i+j=n}k_b+k_d(n-1)p_{n-1}-k_dnp_n,
\end{equation}
where the first two terms on the right-hand side describe the fragmentation process while the last two terms represent the growth of the chain through cell division.
$\sum_{m>n} k_b p_{m}$ corresponds to the rate of obtaining an $n$-mer through the fragmentation of a longer chain. Though there are two possible rupture points (at $n$ and $m-n$) when an $m$-mer breaks to form an $n$-mer, the random selection of one daughter chain causes the prefactors to cancel out. In the second term, $\sum_{i+j=n}k_b = (n-1)k_b$ describes the rupture of the $n$-mer chain into two daughter fragments of lengths $(i,j)$, where $(n-1)$ is exactly the number of connections within an $n$-mer chain. The two terms for the growth process give the rates that an $(n-1)$-mer grows into an $n$-mer, and an $n$-mer grows into an $(n+1)$-mer, respectively.
Using generating functions, we find the steady-state solution: %
\begin{equation}\label{eq:ss_solution_all}
    p_n=\frac{\gamma}{(1+\gamma)^n},~~~\textrm{(all-cell growth)}
\end{equation}
where $\gamma \equiv k_b/k_d$ is the ratio of the break rate to the division rate (see Appendix~\ref{app:gen_func_1}).

\begin{figure}
\includegraphics[width=0.48\textwidth]{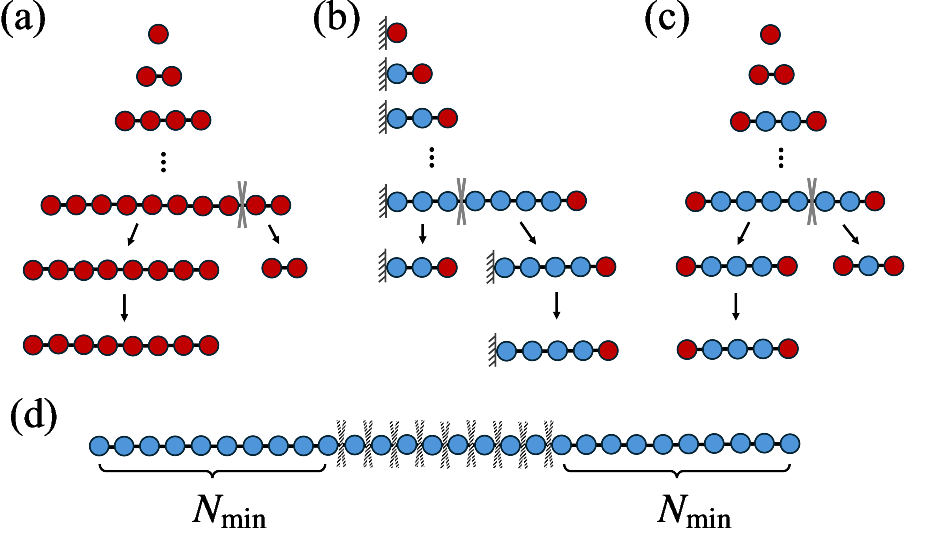}
\caption{Growth and fracture of cell groups under different assumptions. (a) All cells can divide. After a rupture occurs, one daughter chain is selected at random. (b) Only the cell at one end of the chain can divide. (c) Only the cells at the two ends can divide. (d) Potential rupture points when a minimum cluster size $N_\textrm{min}$ is enforced. Rupture can only occur when the chain length $n\geqslant 2N_\textrm{min}$.
Red cells have a non-zero division rate $k_d$, while blue cells have a division rate of zero; gray crosses indicate actual or potential rupture points.}
\label{fig:schematics}
\end{figure}

\begin{figure*}
\includegraphics[width=\textwidth]{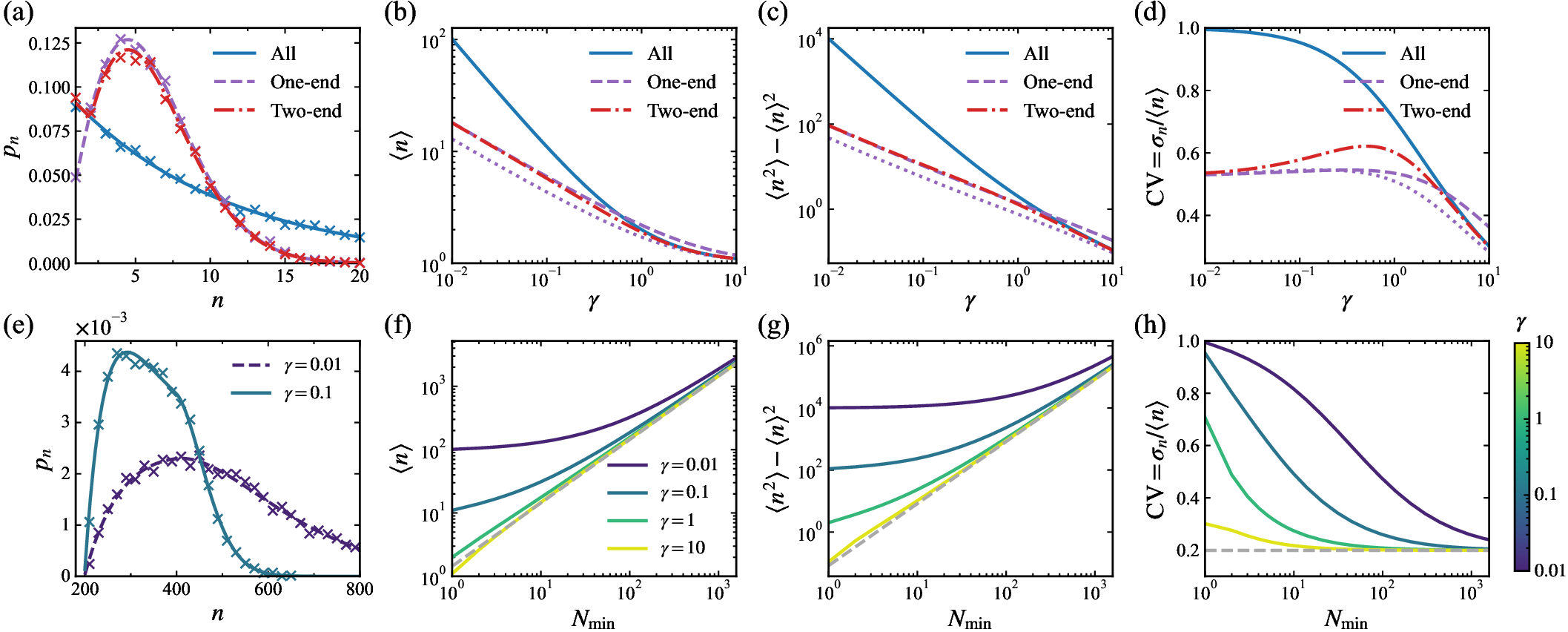}
\caption{(a) Steady-state distribution of chain length $p_n$ for the three growth modes, with $\gamma=0.1$ for the all-cell growth and two-end growth models, and $\gamma$ replaced by $\gamma'
=\gamma/2=0.05$ for the one-end growth model (doubled division rate). Crosses represent simulation results, while lines correspond to theoretical predictions.
Panels (b)--(d) display the mean $\langle n\rangle$, variance $\langle n^2\rangle-\langle n\rangle^2$, and coefficient of variation $\mathrm{CV}=\sigma_n/\langle n\rangle$, where $\sigma_n=\sqrt{\langle n^2\rangle-\langle n\rangle^2}$. Purple dashed lines are the one-end result after doubling the division rate [$\gamma \to \gamma'=\gamma/2$ in Eq.~\eqref{eq:ss_solution_one}], while the purple dotted lines show Eq.~\eqref{eq:ss_solution_one}.
(e) Steady-state distribution $p_n$ when minimum cluster size $N_{\min}=200$. Crosses are simulation results, lines are theoretical predictions by the matrix method.
Panels (f)--(h) show how mean, variance, and $\mathrm{CV}$ vary with $N_\mathrm{min}$ in the all-cell growth model. Gray dashed lines show predictions assuming log-uniform distributions [Eq. \eqref{eq:loguniform}].} %
\label{fig:statistics}
\end{figure*}

So far, we have assumed that all cells in the chain can divide. However,  cell division may be regulated by mechanical and spatial constraints. Cells with fewer neighbors or experiencing lower compression are more likely to divide---``contact inhibition of proliferation''~\cite{puliafito2012collective}. As an extreme limit of contact inhibition, we develop alternate models where only cells at the chain ends divide.
First, consider the case where only one end of the chain grows---e.g., if the other end is constrained by a barrier. In contrast to the all-cell growth model, where the total growth rate scales as $k_dn$ (proportional to chain length), in one-end growth, the total growth rate is fixed at $k_d$  [Fig.~\ref{fig:schematics}(b)]. Here, the master equation becomes
\begin{equation}\label{eq:master_one}
    \frac{\mathrm{d} p_n}{\mathrm{d} t}=\sum_{m>n} k_b p_{m} -p_n\sum_{i+j=n}k_b+k_dp_{n-1}-k_dp_n.
\end{equation}
We find that the steady-state solution is
\begin{equation}\label{eq:ss_solution_one}
    p_n={n\gamma^{1-n}}\Big/{\left(\dfrac{1+\gamma}{\gamma}\right)_n},~~~\textrm{(one-end growth)}
\end{equation}
where $(x)_n=x(x+1)(x+2)\cdots(x+n-1)$ is the Pochhammer symbol (see Appendix~\ref{app:gen_func_2}).
When $\gamma\ll 1$, $p_n$ can be approximated by a Rayleigh distribution 
\begin{equation}
  f(n;\sigma)=n\gamma \mathrm{e}^{-\gamma n^2/2},  
\end{equation}
 with mode $\sigma =1/\sqrt{\gamma}$.

If the chain can grow from both ends, i.e., only the two cells at the two ends can divide, the total growth rate is $2k_d$, except when $n=1$ (the chain has only one particle) where the total growth rate is $k_d$ [Fig.~\ref{fig:schematics}(c)]. %
For this case, the steady-state solution is (see Appendix~\ref{app:gen_func_3})
\begin{equation}\label{eq:ss_solution_two}
    p_n = 
    \begin{cases}
        \gamma/(1+\gamma)^n, & n=1,2,\\
        \dfrac{2+\gamma}{2(1+\gamma)}\dfrac{n(\gamma/2)^{1-n}}{\bm{(}(2+\gamma)/\gamma\bm{)}_n}, & n>2.
    \end{cases}~~~\textrm{(two-end)}
\end{equation}
In the limit of rare breaking $\gamma\ll 1$ when clusters almost always have two ends, $p_n$ converges to Eq.~\eqref{eq:ss_solution_one} with a doubled division rate.

Figure~\ref{fig:statistics}(a) shows our theoretical predictions $p_n$ for the three growth models, validated by Monte Carlo simulations of chain evolution (see Appendix~\ref{app:rfKMC}).
{If all cells can divide, the cluster size distribution is extremely broad with a peak at $n = 1$; if only end cells divide, we have a better-defined peak at $n \approx 1/\sqrt{\gamma}$.}

Larger growth rates relative to break rates lead to larger clusters, but does this also result in greater variability of cluster size?
With the distributions $p_n$, we can compute the means and variances of $n$ [Figs.~\ref{fig:statistics}(b) and \ref{fig:statistics}(c)].
In the limit $\gamma\to 0$, fragmentation is rare, leading to both increased growth and increased variability in all models, while in the limit $\gamma \gg 1$, fragmentation dominates, and clusters become single-celled. %
To quantify the relative variability, we plot the ratio of the standard deviation of cluster size to the mean (coefficient of variation $\mathrm{CV}$) in Fig.~\ref{fig:statistics}(d). We see a key difference between the end-growth models and the all-growth model when $\gamma\ll 1$ and cluster sizes are large---$\mathrm{CV}$ approaches $1$ for the all-growth model, while it approaches $\sqrt{(4-\pi)/\pi}\approx 0.523$ for end-growth. The $\mathrm{CV}$ for end-growth is roughly half that for all-growth, indicating better size control. {If the interior cells can still divide, but at a lower rate, the $\mathrm{CV}$ is intermediate between end- and all-growth (see Appendix~\ref{app:basal}).}

We have so far assumed all rupture rates to be identical, but secreted factors~\cite{vennettilli2022autologous} or stress~\cite{trepat2009physical} may vary across the cluster. {For instance, it is well established that in large epithelial sheets, stress is peaked at the center of the tissue~\cite{trepat2009physical}---suggesting that fractures are more likely to happen at the center of the cluster. This stress profile is not something that would not emerge from our simple model, which creates distributions of cell-cell distances that are uniform across the cluster; to recapitulate this stress profile requires extensive modeling of mechanisms like contact inhibition of locomotion (CIL)~\cite{zimmermann2016contact}. Instead of explicitly modeling the added stress profile from CIL and other factors, we include a somewhat \textit{ad hoc} extension of the model, making clusters more likely to break in the center by assuming that only the junctions sufficiently far from the cluster edge break. This is equivalent to introducing a minimum cluster size $N_\textrm{min}$ [Fig.~\ref{fig:schematics}(d)]. This assumption is a very strong constraint on where breaks can happen---we think of this as the natural opposite limit to our earlier model where breaks are equally likely to occur in all locations along the chain of cells.}
Once we include the constraint that all clusters are larger than $N_\textrm{min}$, we cannot solve the master equation analytically. However, in the steady state, we can write it simply as a matrix equation 
\begin{equation}
    \bm{K}\mathbf{p}=\mathbf{0},
\end{equation}
 which can be easily solved numerically to find $\mathbf{p}$ (see Appendix~\ref{app:matrix}). %
This method leads to good agreement with stochastic simulations [Fig.~\ref{fig:statistics}(e)].%

In the absence of a minimum cluster size, large cluster sizes always come with an unavoidable variability [Figs.~\ref{fig:statistics}(b)--\ref{fig:statistics}(d)], while large rupture rates $\gamma \gg 1$ lead to sizes being precise but $\langle n\rangle \to 1$. %
Given a minimum cluster size, it's possible to have precise size control and finite clusters  [Figs.~\ref{fig:statistics}(e)--\ref{fig:statistics}(h)]. In the limit $\gamma \gg 1$, cell clusters grow from size $N_\textrm{min}$ to $2 N_\textrm{min}$, then fragment in half. This is very similar to the dynamics of a single cell's growth, akin to a sizer model~\cite{ocal2024universal,amir2014cell}, and leads to distributions $p_n$ [Fig.~\ref{fig:statistics}(e)] like those of single cell size distributions~\cite{ocal2024universal, jia2022characterizing,amir2014cell,taheri2015cell}. %
During each growth cycle $t\in[0,t_c]$, we have $\langle n\rangle = N_\textrm{min}\mathrm{e}^{k_d t}$ if all cells can divide, where $t_c = \ln 2/k_d$ is the time required to double the chain's length. Over a long trajectory, any moment within the chain's lifecycle is equally likely to be sampled~\cite{ocal2024universal}. If $t$ is uniformly distributed in one cycle $[0,t_c]$, then $p_n$ should obey a log-uniform distribution, as shown in Fig.~\ref{fig:sm_distrib_all_models}(e):
\begin{equation}
    p_n = \frac{1}{n\ln 2},~~~n\in[N_\textrm{min}, 2N_\textrm{min}], \label{eq:loguniform}
\end{equation}
where the mean is $\langle n\rangle = N_\textrm{min}/\ln 2$ and the variance is $\langle n^2\rangle - \langle n\rangle^2 = (3\ln 2 - 2)\langle n\rangle^2/2$, i.e., the coefficient of variation $\mathrm{CV}$ is independent of $N_\textrm{min}$, as shown in Figs.~\ref{fig:statistics}(f)--\ref{fig:statistics}(h). Similarly, if only edge cells can divide, the chain length increases linearly with time. $p_n$ is then uniformly distributed in $[N_\textrm{min}, 2N_\textrm{min}]$, leading to a mean of $3N_\textrm{min}/2$ and a variance of $N^2_\textrm{min}/12$. When $\gamma\gg 1$, we see mean cluster size scales as $ N_\textrm{min}$ and the variance scales as $N^2_\textrm{min}$, but the $\mathrm{CV}$ goes to a nonzero constant as $\gamma \gg 1$, and is smaller than for clusters with no minimum size. {Interestingly, even given this strict cluster size constraint, the coefficient of variation is still not so small, reaching a minimum of $\sqrt{(3 \ln 2 -2)/2} \approx 0.2$ for the all-growth model. This extreme example shows that the decrease in CV from $1$ to $\approx 0.523$ between all-growth and end-growth is a very relevant change, and even the extremely precise control implied by the minimum cluster size cannot reduce the CV much further.}

{Our example of the minimum cluster size is more motivated by large epithelial tissues, where the role of CIL and spatially-varying traction forces are better established. This is why we focus on large $N_\textrm{min}$ in Fig.~\ref{fig:statistics}. However a minimum cluster size does reduce variability even if the minimum is quite small---compare finite $N_\textrm{min}$ in Fig.~\ref{fig:statistics}(h) with no minimum ($N_\textrm{min} = 1$).}

{We note, however, that the strict minimum cluster size is a vast oversimplification, and it is likely that even if fracture is much more likely at the center, clusters likely cannot achieve the tight control limit of Figs.~\ref{fig:statistics}(e)--\ref{fig:statistics}(h) when $\gamma$ becomes large. If we weaken the strict assumption of a minimum cluster size by assuming a basal break rate $k_b^0$ for all junctions, even a small $k_b^0$ creates a relevant population of small clusters (see Appendix~\ref{app:basal}), increasing the variability. We show in Appendix~\ref{app:basal}) how this basal rate can interpolate between the limit of all junctions having identical break rates and only a few junctions having nonzero break rates---spanning a large range of possible control strategies.}

\subsection{Combining mechanics and statistics}\label{sec:combine}
According to our theory, the break rate $k_b$ depends on four key parameters related to cell motility and mechanical properties of the cluster---$\tau$, $\mu k$, $D$, and $\delta \ell_b$.
While these parameters have not yet been measured simultaneously for a single cell type, we can provide reasonable estimates.
Experiments measuring velocity correlations of HaCaT cells~\cite{selmeczi2005cell} give a persistence time $\tau$ of roughly $0.5$ hour and $D$ of around $\sim 100\,\textrm{\textmu}\mathrm{m^2/h}$, leading to an active velocity $v_n$ on the order of $\sqrt{D/\tau}\approx 15\,\textrm{\textmu}\mathrm{m/h}$.
The mobility $\mu$ and spring constant $k$ together create a relaxation timescale $1/\mu k$, estimated to be approximately $15\,\mathrm{min}$ based on measurements for MDCK cells~\cite{jain2020role} (details in Appendix~\ref{app:AOUP_params}).
According to  Eq.~\eqref{eq:break_rate}, the break rate $k_b$ is highly sensitive to the critical break length $\delta\ell_b$.
{As a rough guess, $\delta\ell_b=5\,\textrm{\textmu}\mathrm{m}$ yields a mean escape time $\tau_\mathrm{esc}\approx 17.6\,\mathrm{h}$, or a break rate of approximately $0.057\,\mathrm{h^{-1}}$. This is the right order of magnitude for experiments in germline cyst fracture~\cite{levy2024tug} and cancer cell dissociation~\cite{law2023cytokinesis,wang2025confinement}. We use these values as our default parameters (Table~\ref{tab:param}).} %

\begin{figure}
\hspace{-0.2cm}
\includegraphics[width=0.48\textwidth]{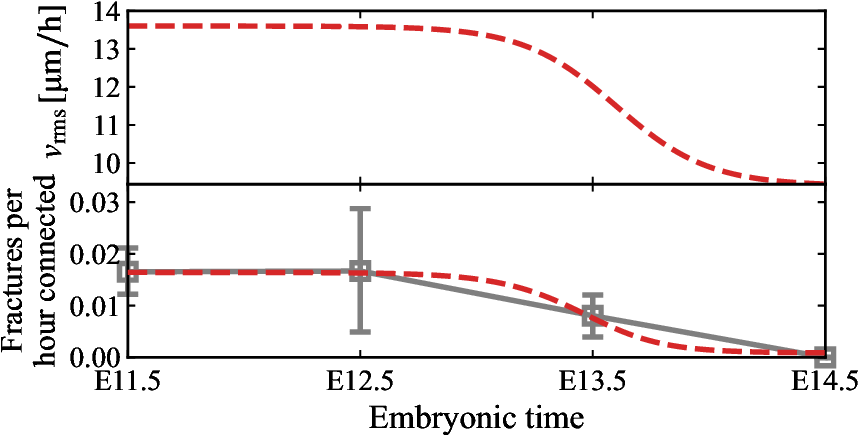}
\caption{{A modest change in cell motility can account for the observed decrease in fracture rates over embryonic time. The upper panel shows a logistic decrease in the typical cell speed, $v_\textrm{rms}=\sqrt{D/\tau}$ which we hypothesize underlies the fracture rate reduction. In the lower panel, symbols with error bars are experimental data from Fig.~4C of Ref.~\cite{levy2024tug}; the red dashed line shows the corresponding theoretical predictions for the break rate $k_b$, calculated from the upper panel's $v_\textrm{rms}$ using Eq.~\eqref{eq:break_rate}.}}
\label{fig:expt}
\end{figure}

Our results suggest that clusters of cells may regulate their size not only by changing division rate but also by changing cell motility or persistence or cell-cell adhesion. Recent experiments found that germline cells have motility that decreases as cysts develop, decreasing fracture over time and controlling cyst size~\cite{levy2024tug}. How much does a change in cell properties change cluster size?
{In Ref.~\cite{levy2024tug}, Levy \textit{et al.} measured the break rates as the \textit{fractures per hour connected}, defined as the number of fracture events observed per hour of intact intercellular bridges, for germline cysts at different embryonic stages (lower panel, Fig.~\ref{fig:expt}). They also reported that germline cysts have a cell cycle length of $\sim16$ hours.
Thus, the break rate and division rate of germline cysts are approximately $0.02\,\mathrm{h^{-1}}$ and $\ln2/16\approx0.04\,\mathrm{h^{-1}}$,} i.e., the ratio $\gamma\approx 0.5$, corresponding to a mean cluster size $\langle n\rangle\approx 3$
in the all-cell growth model.
(We note this is not the value found by Ref.~\cite{levy2024tug}---their germline cysts do not reach the steady state. If we simulate for only $5$ division times, we get similar results to their experiment and model; see Appendix~\ref{app:rfKMC}.)
For germline cysts, we set $\delta \ell_b = 6.5\,\textrm{\textmu}\mathrm{m}$ to match  $k_b$ to experiment.
Cell motility and cell-cell adhesion change $k_b$ and thus change $\gamma$, changing cluster size distributions.
For instance, if cells double their adhesiveness by increasing $k$, $\gamma$ decreases to approximately $0.0067$, leading to a larger cluster size $\langle n\rangle\approx 150$ with significantly higher variability $\sigma_n\approx 150$.
At long developmental times, the break rate of germline cysts drops to under $1/190\,\mathrm{h^{-1}}$, which Ref.~\cite{levy2024tug} attributes to changes in cell motility. We find this drop does not require a dramatic change in cell speed. To decrease the break rate to this level, $D$ must decrease by more than a factor of $1.5$, i.e., the typical speed of cells needs to decrease by more than a factor of $1.2${, as shown in Fig.~\ref{fig:expt}, where a small decrease in $v_\textrm{rms}=\sqrt{D/\tau}$ captures the observed drop in break rates. We note that this $v_\textrm{rms}$, while consistent with the observed slowing of the cells' motion, is only hypothesized; extracting parameters like $D$ and $\tau$ directly from the experiment is nontrivial and requires additional imaging beyond that done in Ref.~\cite{levy2024tug}.}

{
\subsection{Survival probability has a surprising universal form for  all-cell growth}\label{sec:survival}
}

{At steady state, how long does a given cluster stay intact before it fractures into two pieces? Our initial intuition for this survival problem was that survival time should depend strongly on the break rate $k_b$ of an individual junction. However, the break rate $k_b$ and division rate $k_d$ affect this time scale in several ways, making their effects not immediately obvious. For instance, a larger $k_b$ makes junctions more likely to break, but it also shifts the steady-state distribution toward smaller clusters, which are less likely to fracture---simply because they contain fewer cell-cell junctions.
To tackle the questions, we compute the survival probability, i.e., the probability that the time $T$ of the first fracture event is larger than the current time, $S(t)=\mathbb{P}(T>t)$, for a given cluster {which we draw from the steady-state distribution of clusters at $t = 0$. (We used similar approaches to quantify experimental data on cancer cell rupture in \cite{law2023cytokinesis,wang2025confinement}.)}
Our Monte Carlo simulation results reveal that, for the all-cell growth model, the survival probability follows an exponential form:
\begin{equation}
    S(t)=\mathrm{e}^{-k_d t},
\end{equation}
which is surprisingly independent of the break rate $k_b$, and depends solely on the division rate $k_d$, as shown in Figs.~\ref{fig:survival_probs}(a) and \ref{fig:survival_probs}(b).}

{
We can show that this form of the survival probability holds not just for our simplest model, but a broader generalization of the all-cell growth model [Eq.~\eqref{eq:master_all}]:
\begin{equation}\label{eq:master_survival}
    \frac{\mathrm{d}p_n}{\mathrm{d}t}=\sum_{m>n}q_{mn} p_m - \lambda(n) p_n  + (n-1)k_d p_{n-1} - nk_d p_n,
\end{equation}
where $q_{mn}$ is the transition rate at which an $m$-cell cluster breaks into an $n$-cell cluster. {Our earlier all-cell growth model is the simple case where $q_{mn}$ is just set to a constant $k_b$.}
The transition rate is symmetric in fragment size: $q_{mn}=q_{m,m-n}$. The total ``hazard rate'' for an $m$-cell cluster $\lambda(m)$---{i.e., the total rate of fragmentation for a cluster of size $m$ to \textit{any} size}---is then given by $\lambda(m) = \sum_{n\geqslant 1}^{m-1}q_{mn}$.
Now, let us consider a more general class than our earlier all-cell growth model, in which the transition rate depends on the pre-fracture cluster size $m$, but not on the post-fracture fragment size $n$:
\begin{equation}\label{eq:survival_assm}
    q_{mn}=\mathcal{Q}(m).
\end{equation}
This would include fracture rates in models where, for instance, each bond in a larger cluster is less likely to break. Given Eq.~\eqref{eq:survival_assm},
the total hazard rate for an $n$-cell cluster becomes $\lambda(n)=\sum_{i+j=n}\mathcal{Q}(n)=(n-1)\mathcal{Q}(n)$, since there are $n-1$ possible ways to partition an $n$-cell cluster into two subclusters. In the steady state, we find a recurrence relation between successive probabilities:
\begin{equation}\label{eq:recurrence}
    \frac{p_n}{p_{n-1}}=\frac{(n-1)k_d}{(n-1)k_d + \lambda(n)}=\frac{k_d}{k_d+\mathcal{Q}(n)}.
\end{equation}
As this is a generalization of the all-cell growth model, the solution in Eq.~\eqref{eq:ss_solution_all} automatically satisfies the recurrence relation.
}

\begin{figure}
\includegraphics[width=0.48\textwidth]{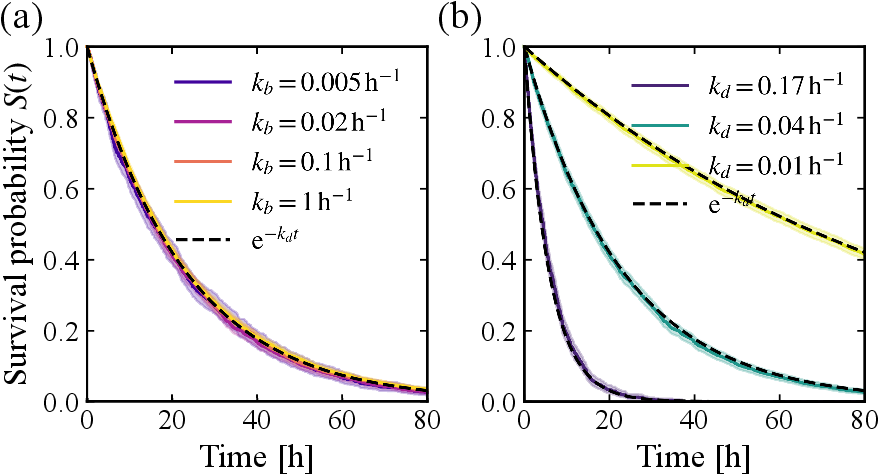}
\caption{{Dependence of survival probability $S(t)$ on break and division rates in the one-dimensional all-cell growth model.
(a) Survival curves for varying break rates $k_b$, with fixed division rate $k_d$, all collapse onto the exponential decay $S(t)=\mathrm{e}^{-k_d t}$ (dashed line).
(b) Survival probabilities $S(t)$ for three different division rates $k_d$, corresponding to  cell cycles of $4, 16$, and $64$ hours, respectively.}
{We apply standard Kaplan-Meier survival analysis~\cite{kaplan1958nonparametric} to the data from our rejection-free kinetic Monte Carlo simulations; error bars represent 95\% confidence intervals.}
}
\label{fig:survival_probs}
\end{figure}

{
To compute the total survival probability $S(t)$, we first consider the conditional survival probability $u_n(t)=\mathbb{P}(T>t|N(0)=n)$, defined as the probability that no fracture occurs by time $t$ given an initial chain length $N(0)=n$, where $N(t)$ denotes the chain length at time $t$. 
{
For a Markov process with a constant hazard rate $\lambda$, the probability that no failure occurs over a time interval $\Delta t$ is simply given by $\mathrm{e}^{-\lambda\Delta t}$. If the rate $\lambda(t)$ varies with time, this probability can be generalized with an integral $\exp\left(-\int_0^{\Delta t}\lambda(t')\mathrm{d}t'\right)$~\cite{kalbfleisch2002statistical}. In our case, the total hazard rate $\lambda(t)=\lambda\bm{(}N(t)\bm{)}$ depends on a stochastic process $N(t)$, so we must average over all possible trajectories of $N(t)$:} %
\begin{equation}
\label{eq:conditional_average}
u_n(t)=\mathbb{E}^N\left[\left.\exp\left(-\int_0^t\lambda(t')\mathrm{d}t'\right)\right|N(0)=n\right],
\end{equation}
where $\mathbb{E}^N[\cdot]$ represents the expectation over all realizations of the stochastic process $N(t)$, which can be expressed as a path integral $\mathbb{E}^N[\cdot]=\int\cdot\mathbb{P}[N]\mathcal{D}N$.}

{We can work out the average in Eq.~\eqref{eq:conditional_average} using the Kolmogorov backward equation (KBE) describing the transition between different cluster sizes $N$. For such a discrete-state, continuous-time Markov jump process, the KBE is~\cite{ross1995stochastic, Norris_1997}:
\begin{equation}
    \partial_tP_{ij}(t)=\sum_k w_{ik} P_{kj}(t),
\end{equation}
where $P_{ij}(t)=\mathbb{P}(N(s+t)=j|N(s)=i)$ represents the probability that the system in state $i$ at time $s$ will be in state $j$ at a time $t$ later, and $w_{ij}$ is the transition rate from state $i$ to state $j$, with $w_{ii} = -\sum_{j \neq i} w_{ij}$.
To compute the survival time, we are interested in the dynamics of the cluster to reach an absorbing state of ``ruptured'', which we denote as the ``0'' state.  
The clusters that have ruptured are in the ``$0$'' state---so $1-P_{n0}(t)$ is the conditional survival probability $u_n(t)$, and we can write $\partial_t u_n = -\partial_t P_{n0} = -\sum_{k} w_{nk}P_{k0} = - \sum_{k} w_{nk} (1-u_k) = \sum_k w_{nk} u_k$, because $\sum_k w_{nk} = 0$. This means the KBE implies
\begin{equation}
    \partial_t u_n(t)=\sum_{k> 0} w_{nk}u_k(t),
\end{equation}
where we have removed $k = 0$ because $u_0 = 0$ because $P_{00}=1$ for the absorbing state. 
We separate the rates of the processes in Eq.~\eqref{eq:master_survival} into those that involve a rupture (a total rate $\lambda(n)$ from $n$ to any ruptured condition) and those that don't (division). The rate of any fracture from the state with $n$ cells is $w_{n0}=\lambda(n)$. Fracture is absorbing in this survival calculation, so $w_{0j}=0$ for $j \neq 0$. The rate of the cluster growing from $n$ to $n+1$ via division is $w^\textrm{division}_{n,n+1} = nk_d$. From $w_{nn} = -\sum_{k\neq n} w_{nk}$, we see $w_{nn} =-w_{n,n+1}-w_{n0}= -n k_d -\lambda(n)$ for $n \neq 0$. 
Including all these rates, we find that}
{
\begin{equation}\label{eq:KBE}
    \partial_t u_n =-\lambda(n) u_n + n k_d (u_{n+1} - u_n).
\end{equation}
The total survival probability is then obtained by averaging $u_n(t)$ over the distribution of initial chain lengths.}
{
If we assume that the system is at steady state, it is then reasonable to use the steady-state $p_n$ as the distribution of initial chain lengths, and we therefore write}
{
\begin{equation}
    S(t)=\sum_{n\geqslant 1}\mathbb{P}\bm{(}N(0)=n\bm{)}\cdot u_n(t)=\sum_{n\geqslant 1} p_{n}\cdot u_n(t).
\end{equation}
By computing the time derivative of $S(t)$, we have $\dot{S}(t)=\sum_{n\geqslant 1} p_n \dot{u}_n$, and substituting Eq.~\eqref{eq:KBE} and the recurrence relation [Eq.~\eqref{eq:recurrence}] into it,  we obtain
\begin{equation}
    \dot{S}(t)=-k_d\sum_{n\geqslant 1} p_n u_n =-k_d S(t).
\end{equation}
Therefore, we arrive at the striking result that the survival probability decays exponentially as $S(t)=\mathrm{e}^{-k_d t}$ independent of the break rate $k_b$, and determined solely by the division rate $k_d$.
We note that this derivation is independent of dimensionality and relies only on the assumption in Eq.~\eqref{eq:survival_assm} and the all-cell growth mechanism. In our one-dimensional all-cell growth model, where $q_{mn}=k_b$ is constant, the derived $S(t)=\mathrm{e}^{-k_d t}$ is thus the exact solution, as confirmed by the Monte Carlo simulation results shown in Fig.~\ref{fig:survival_probs}.
}

{The intriguing result that the survival probability depends only on $k_d$ depends on the assumption of steady state. However, germline cysts do not reach steady state---so for a comparison to experiment, an ensemble starting from a fixed number of cells corresponding to the initially observed condition would be easier to observe and compare to experiment. The quantity $u_1(t)$---the survival probability conditional on starting from a single cell---might be of particular interest. We present the analytical expression for the conditional survival probability $u_n(t)$ in our one-dimensional all-cell growth model in Appendix~\ref{app:conditional_survival}.}

{
An essential assumption in this derivation is the all-cell growth mechanism, which ensures the simple structure of the corresponding Kolmogorov backward equation [Eq.~\eqref{eq:KBE}]. If we choose alternative growth mechanisms---such as edge-growth, we find that the $S(t)$ curve becomes dependent on the break rate $k_b$ [Fig.~\ref{fig:suppl_survival}(a)], and the exponential form $\mathrm{e}^{-k_d t}$ now serves as a lower bound in the limit of large $\gamma=k_b/k_d$, where different growth models converge [see Figs.~\ref{fig:statistics}(b)--\ref{fig:statistics}(d)]. Additionally, imposing a minimum cluster size $N_\textrm{min}$---which breaks our assumption of fracture rates $q_{mn}$ being independent of $n$---also violates the universal $\mathrm{e}^{-k_d t}$ decay [Fig.~\ref{fig:suppl_survival}(b)]. In this case, $\mathrm{e}^{-k_d t}$ will serve as an upper bound for the survival curves at various $k_b$, since smaller $\gamma$ values allow clusters to grow significantly larger than $2N_\textrm{min}$, letting them to fracture similarly to those in the standard all-cell growth model.}

\begin{figure*}
\includegraphics[width=\textwidth]{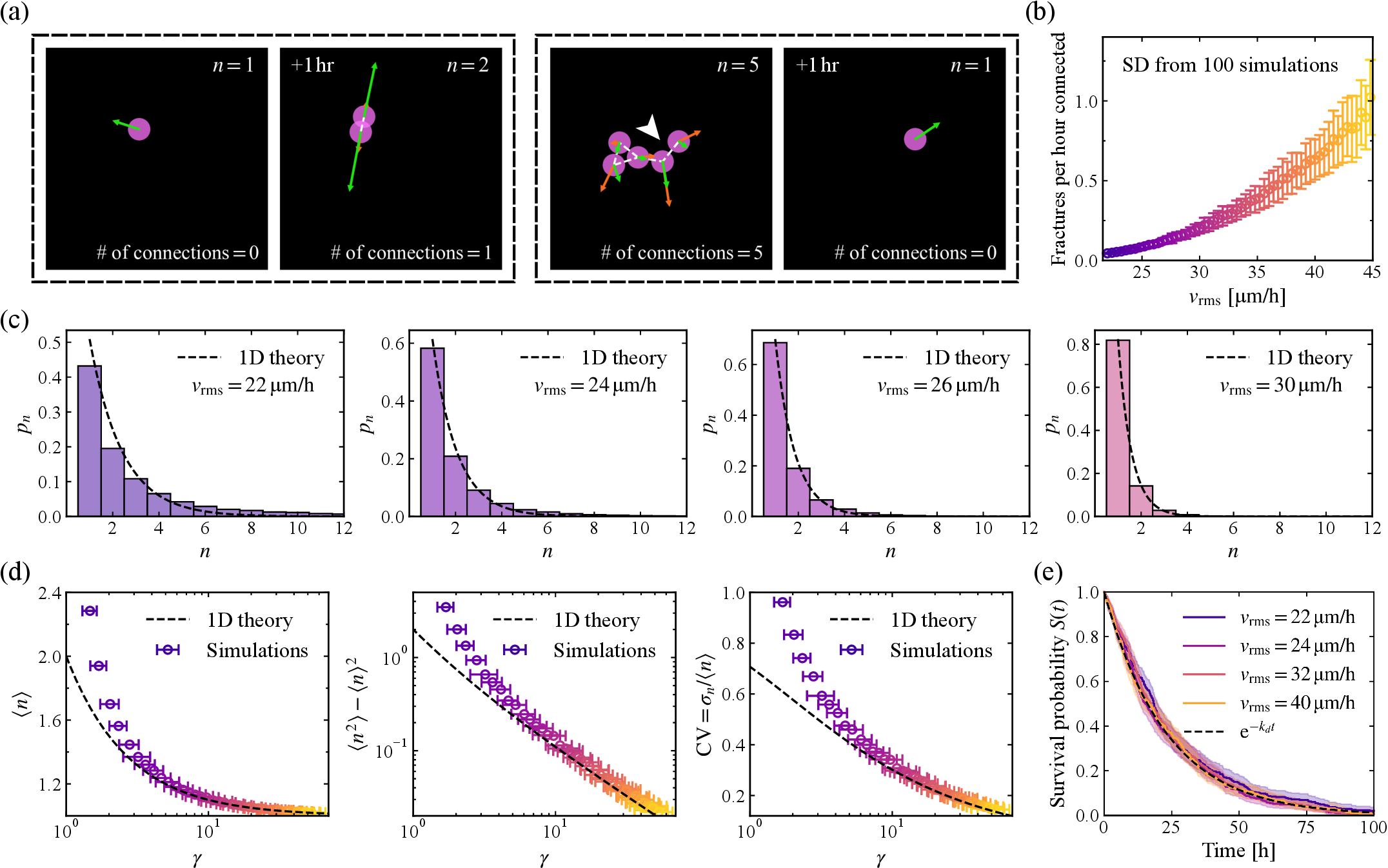}
\caption{{2D simulation results.
(a) Representative snapshots from a simulation with $v_\textrm{rms}=22\,\textrm{\textmu}\mathrm{m/h}$. The first dashed box shows a cell division event; the second dashed box shows a fracture event (highlighted by the white arrow), where only one of the two resulting sub-clusters is retained. Green arrows indicate instantaneous cell velocities, orange arrows indicate active self-propulsion velocity, and white dashed lines denote intercellular junctions.
(b) Fractures per hour connected increase with the root-mean-square cell speed $v_\textrm{rms}$.
(c) Lineage size distributions $p_n$ for different values of $v_\textrm{rms}$, compared to the 1D theoretical prediction $\gamma/(1+\gamma)^n$ (dashed lines), where we have used the corresponding fractures per hour connected in panel (b) as the break rate $k_b$ to compute $\gamma=k_b/k_d$.
(d) How the mean, variance, and CV of cluster size vary with $\gamma$.
(e) Survival probabilities for a wide range of $v_\textrm{rms}$ collapse to $S(t)=\mathrm{e}^{-k_d t}$.
{Kaplan-Meier analysis is used to generate survival curves, and 95\% confidence intervals are shown as error bars.}
Division rate in the simulations is fixed at $k_d=(\ln 2/16)\,\mathrm{h^{-1}}$.}}
\label{fig:2d_simulations}
\end{figure*}

{
\section{2D simulations}\label{sec:2d}
}

{
Most of the preceding results are derived in a one-dimensional geometry, which simplifies analysis but does not capture the spatial complexity of most biological tissues---except for special cases such as germline cysts or tissues under strong confinement. Fracture of a two- or three-dimensional tissue is fundamentally different from our 1D model, because the fracture must cross the entire tissue, rather than break a single link. What results from the simple 1D model can be generalized to a more realistic setting? We extend our model to two dimensions using simulations. Specifically, we implement a two-dimensional model treating cells as active Ornstein-Uhlenbeck (OU) particles, governed by the equations:
\begin{subequations}
\begin{eqnarray}
    \dot{\mathbf{x}}_n &=& -\mu \bm{\nabla}\Phi + \mathbf{v}_n(t),\\
    \tau \dot{\mathbf{v}}_n&=&-\mathbf{v}_n+\sqrt{2D}\bm{\xi}_n(t),
\end{eqnarray}
\end{subequations}
where the Gaussian noises satisfy $\langle \xi_{n\alpha}(t)\xi_{n'\beta}(t')\rangle = \delta_{nn'}\delta_{\alpha\beta}\delta(t-t')$, with Greek indices corresponding to spatial components.
{As in 1D, we have a springlike potential between cells, $\Phi=\frac{1}{2} k\sum_{(i,j)}(|\mathbf{x}_i -\mathbf{x}_j|-\ell_0)^2$, where $(i,j)$ denotes the summation over all connected pairs.}
In this 2D setting, the root-mean-square speed of an isolated cell is given by
\begin{equation}
    v_\textrm{rms}=\sqrt{2D/\tau}.
\end{equation}
Upon division, the original cell is replaced by a mother-daughter pair positioned at the same location~\cite{soumya2015coherent}, with a small displacement $2\ell_0/3$ between them introduced along a random direction. 
The two daughter cells are initially given velocities {with magnitude $v_\textrm{rms}$} pointing away from the cell-cell contact along this displacement vector.
Intercellular connections break when stretched beyond the threshold length $\ell_0+\delta\ell_b$,
and we allow connections to be re-established when the distance between two cells becomes $\leqslant\ell_0+\epsilon$, where $\epsilon = 10^{-3}\ell_0$ in our simulations. {When we find that a cluster has broken into two separate pieces (established by finding the connected components of the cluster), we randomly select one of the two clusters as the daughter cluster we are following.}
Figure~\ref{fig:2d_simulations}(a) shows representative snapshots of a cell division event and a cluster undergoing fracture and a daughter cluster being selected (indicated by the white arrow).
}

{
To explore the effect of cell motility, we vary the typical cell speed $v_\textrm{rms}$ by tuning $D$ while keeping $\tau$ fixed.
As shown in Fig.~\ref{fig:2d_simulations}(b), the mean number of fractures (breakage of cell-cell connections) per hour connected increases monotonically with $v_\textrm{rms}$.
We then use this fracture rate from simulations as the break rate $k_b$; higher cell speeds $v_\textrm{rms}$ result in larger values of $k_b$, and consequently, a larger break-to-division ratio $\gamma=k_b/k_d$.
By measuring the lineage cluster size distribution from simulations, we find that, for the all-cell growth model, the one-dimensional theoretical prediction $p_n=\gamma/(1+\gamma)^n$ [Eq.~\eqref{eq:ss_solution_all}] can still describe the 2D results in a wide range of cell motility, as shown in Fig.~\ref{fig:2d_simulations}(c).
Noticeable deviations from the theoretical distribution $p_n$ occur mainly at small cluster sizes $n$ for low cell speeds.
Further comparison of the mean, variance, and coefficient of variation (CV) of cluster sizes as functions of $\gamma$ [Fig.~\ref{fig:2d_simulations}(d)] confirms that the 1D theory can describe the 2D simulation outcomes, especially in the regime of large $\gamma$ (high motility).
This agreement can be understood by noting that in the large $\gamma$ limit---where cell-cell junctions frequently rupture---only small clusters, such as one- or two-cell units, are likely to persist. As a result, the system effectively behaves as quasi-one-dimensional, and the statistical properties converge to those predicted by the 1D model.
}

{
Interestingly, the exponential survival probability $S(t)=\mathrm{e}^{-k_d t}$ also shows up in our 2D simulation results [Fig.~\ref{fig:2d_simulations}(e)].
{Is this reasonable? The derivation of this result in Sec.~\ref{sec:survival} assumes that the transition rate from clusters of size $m$ to $n$ depends only on $m$. This is more general than our earlier results assuming a strict 1D model, and in principle, this assumption is independent of dimension. However, this choice of assumption is most natural for a generalization of the 1D model---summarizing the clusters solely by size in 2D is potentially suspect given the potential roles of cluster geometry and topology. In fact, the assumption that the fracture rate does not depend on the final cluster size [Eq.~\eqref{eq:survival_assm}] is not strictly satisfied in the 2D simulations.}
We have quantified $q_{mn}$ by counting the number of occurrences where an $m$-cell cluster fragments into an $n$-cell cluster in our simulations. We find that, for a fixed pre-fracture size $m$, the values of $q_{mn}$ are not uniformly distributed across different $n$---the cluster is more likely to break asymmetrically, producing either small or large fragments rather than equal-sized ones, as shown in Fig.~\ref{fig:suppl_survival}(c).
Nevertheless, the exponential form remains valid across a broad range of cell motility in 2D simulations, as shown in Fig.~\ref{fig:2d_simulations}(e), with good agreement observed even at $v_\textrm{rms}=22\,\textrm{\textmu}\mathrm{m/h}$ where the predictions for cluster statistics fail in Fig.~\ref{fig:2d_simulations}(d).
}

{A caveat in our analysis of the 2D system is that we are not using the analytical expression for the break rate [Eq.~\eqref{eq:break_rate}] to directly predict the break rate in our 2D simulations. Instead, we extract the break rate $k_b$ using the measured fractures per hour connected from simulations. This is because Eq.~\eqref{eq:break_rate} is derived in one dimension and assumes the limit $N\to\infty$, making its prediction quantitatively incorrect for the 2D system.
We also note that the model we show here describes a generic two-dimensional cluster, and does not include some features that we think would be appropriate for a full model of the germline cysts. For instance, our cell-cell interactions are non-specific intercellular connections that can break and reform based on intercellular distances---effectively modeling E-cadherin-based adhesion. This differs from germline cysts, where intercellular bridges (ICBs) form specifically between sister cells during division and do not reconnect once severed~\cite{ikami2023branched,levy2024tug}.}

\section{Discussion}\label{sec:discussion}
Our work provides a route to understand quantitatively how the size of groups of cells ranging from cancer to organs to organisms can emerge from balancing growth and random cell motility. Our model shows that differences in where cell division and rupture occur can lead to different characteristic distributions of cluster sizes, and that biologically relevant regulation of cluster size can occur from relatively small changes in cell motility. From a physics standpoint, our results show that even in a very simple one-dimensional active material, the collective rupture of a link is nontrivial and can't be understood from the properties of a single pair of cells---unlike in an equilibrium version of our model. 
This qualitative difference with the simple active trap model~\cite{woillez2020active} suggests that rupture rates may be sensitive to other collective features, e.g., differing between branched and linear chains of cells, or reflecting the degree of cell-cell correlation of velocities.

{
Though the remarkable exponential decay of the survival probability $S(t)=\mathrm{e}^{-k_d t}$, derived in Sec.~\ref{sec:survival}, is independent of dimensionality and break rate $k_b$, and depends solely on the division rate $k_d$, it relies on two important underlying assumptions.
First, the rate of a cluster splitting into two pieces must depend only on the pre-fracture cluster size $m$, and not on the post-fracture fragment size $n$, though the functional dependence on $m$ can be arbitrary.
This first assumption may be able to be relaxed somewhat. Even though in our 2D simulations, clusters of $m = 10$ cells are more likely to have single cells break off than split equally [Fig.~\ref{fig:suppl_survival}(c)], the total survival probability still follows the predicted exponential form $\mathrm{e}^{-k_d t}$ across a wide range of cell motility. This may reflect either that the exponential form holds for a more general form of our nitial assumption, or merely that within the parameter space studied in Fig.~\ref{fig:2d_simulations}(e) that typical cluster sizes are small enough that the asymmetries in $q_{mn}$ are not relevant. {If this first assumption is more dramatically violated, e.g., with a minimum cluster size, $S(t)$ will depend on the break rate $k_b$ [Fig.~\ref{fig:suppl_survival}(b)].}
The second key assumption is the all-cell growth mechanism. When alternative growth mechanisms are considered, the universal exponential form no longer holds [Fig.~\ref{fig:suppl_survival}(a)]. In this sense, the connection between the survival probability $S(t)$ and the single-cell division rate is a signature of simple all-cell growth.}

We have generally chosen analytic tractability over biological detail in our approach to capture the key elements of this problem. {There are many possible features that could be included in more general models of size control via fracture. These include the presence of polymer networks within the cell which can alter its ability to load stress onto the cell-cell junction~\cite{duque2024rupture}, cell-matrix interactions~\cite{zhang2025emergence,perez2024deposited}, cell shape~\cite{wang2025confinement,chen2022activation} and its coupling to division~\cite{kaiyrbekov2023migration,puliafito2012collective}, or mechanosensitive feedback~\cite{mcevoy2022feedback}, as well as cell-cell interactions like contact inhibition of locomotion~\cite{mayor2010keeping, camley2014polarity, wang2024limits,zimmermann2016contact} and collective alignment of cell polarity~\cite{camley2017physical,jain2020role}.} %
Determining to what extent these features only change the relevant energy barrier, rescaling $k$, or qualitatively change our picture is an important open question.
~\\

\begin{acknowledgments}
The authors acknowledge support from NIH Grant No.~R35GM142847.
This work was carried out at the Advanced Research Computing at Hopkins (ARCH) core facility, which is supported by the National Science Foundation (NSF) Grant No.~OAC~1920103.
We thank Cody Schimming and Emiliano Perez Ipi\~na for a close reading of the manuscript. We thank Isabella Leite, Stanislav Shvartsman, and Eszter Posfai for useful conversations.
\end{acknowledgments}

\section*{Data availability}
The data and code that support the findings of this article are openly available~\cite{wang2026code}.

\appendix

\section{Rates for breaking chains using Kramers' theory}
\subsection{Results for two particles}\label{app:two_particle}
{The rupture rate of a pair of cells can be found straightforwardly through a very simple generalization of the approach of Ref.~\cite{woillez2020active}, who mapped overdamped active Ornstein-Uhlenbeck particles (AOUPs) in a harmonic well to an equivalent thermal system. Here, we show that the same holds for a pair of harmonically bound AOUPs.}
For a pair of particles $\{x_1, x_2\}$ with a distance $\ell=x_2-x_1$ between them, we have
\begin{eqnarray*}
    \dot{x}_1&=&\mu k\delta\ell + v_1,\\
    \dot{x}_2&=&-\mu k\delta\ell + v_2,
\end{eqnarray*}
where $\delta\ell=\ell-\ell_0$ is the stretched length of the spring.
Thus we can obtain $\dot{\delta\ell}=-2\mu k\delta\ell + \delta v$, where $\delta v = v_2 -v_1$ is the solution of $\tau\dot{\delta v} =-\delta v + \sqrt{4D}\xi(t)$---i.e., $\delta v$ is still an Ornstein-Uhlenbeck process but with correlation $\langle\delta v(t)\delta v(t')\rangle = (2D/\tau)\mathrm{e}^{-|t-t'|/\tau}$. By a standard change of variable $p=-2\mu k\delta\ell +\delta v$, we obtain
a new set of Langevin equations for a thermal Brownian particle with displacement $\delta \ell$~\cite{woillez2020active, woillez2020nonlocal}:
\begin{equation}\label{eq:recast_EOM}
\begin{cases}
    \phantom{\tau}\dot{\delta\ell}=p,\\
    \tau\dot{p}=-\eta p -U' + \sqrt{4D}\xi(t),
\end{cases}
\end{equation}
where the potential $U(\delta\ell)=\mu k\delta\ell^2$ is harmonic, the persistence time $\tau$ is acting as a mass, and the drag $\eta =1+2\mu k\tau$. Thus an effective temperature can be defined by
\begin{equation}
    k_BT_\textrm{eff} \coloneq \frac{2D}{\eta} = \frac{2D}{(1+2\mu k\tau)}.
\end{equation}
Then we can write out the Fokker-Planck equation (aka Klein-Kramers equation) from Eq.~\eqref{eq:recast_EOM}:
\begin{equation*}
    \partial_t P(\delta\ell, p, t)=-p\partial_{\delta\ell} P + \frac{1}{\tau}\partial_p(\eta p + U')P + \frac{2D}{\tau^2}\partial^2_p P,
\end{equation*}
and the stationary solution is given by a Boltzmann-like measure
\begin{equation}
    P_s(\delta\ell, p)=\mathcal{Z}\exp\left(-\frac{U}{k_B T_\textrm{eff}} -\frac{\tau p^2}{2k_B T_\textrm{eff}}\right),
\end{equation}
where $\mathcal{Z}$ is the partition function.
The variance of $\delta\ell$ follows directly,
$$\langle\delta\ell^2\rangle =\frac{k_B T_\textrm{eff}}{U''} = \frac{D}{\mu k(1+2\mu k\tau)}.$$

From standard Kramers' theory, for a trap in the potential $U(\delta\ell)$ of size $\delta\ell_b$, the mean escape time, when the effective temperature is small compared to the potential barrier height $\Delta U =U(\delta\ell_b)-U(0)$, is given by~\cite{gardiner2009stochastic, hanggi1990reaction, haunggi1994colored}
\begin{equation*}
    \tau_\textrm{esc}=\frac{2\pi}{\omega_0}\exp\left(\frac{\Delta U}{k_BT_\textrm{eff}}\right)=\pi\sqrt{\frac{2\tau}{\mu k}}\exp\left[\frac{\mu k\delta\ell_b^2(1+2\mu k\tau)}{2D}\right],
\end{equation*}
where $\omega_0 = \sqrt{U''/\tau}=\sqrt{2\mu k/\tau}$ is the undamped angular frequency of the system. The subexponential correction $\tau_0=2\pi/\omega_0$ applies in the limit of moderate friction but may vary under different friction regimes~\cite{wexler2020dynamics, hanggi1990reaction}.

\subsection{Results for a long chain}\label{app:chain}
We can explicitly write Eq.~\eqref{eq:EOM} in the main text as:
\begin{equation*}
    \dot{x}_n = \mu k (x_{n+1}-2x_n+x_{n-1}) +v_n.
\end{equation*}
Assuming that $x_n(t)=n\ell_0+\delta x_n(t)$, where $\delta x_n$ represents the deviation of $x_n$ around the lattice point $n\ell_0$, we find that $\delta x_n$ obeys the same equation as $x_n$:
\begin{equation}\label{eq:EOM_PBCs}
    \dot{\delta x}_n = \mu k (\delta x_{n+1}-2\delta x_n+\delta x_{n-1}) +v_n.
\end{equation}
For a long chain with $N\gg 1$ particles, it is simpler to solve for the fluctuations of cell-cell distances if we impose periodic boundary conditions, $\delta x_{n+N} = \delta x_{n}$. In principle this could change the amplitude, but we find that in our simulations in Appendix~\ref{app:AOUP_simulation} with free boundaries at the end of the chain, the variance $\langle \delta \ell_n^2\rangle$ does not depend on $n$ when the number of cells $N$ is large.
An alternate approach would be to use a continuum approximation $(\delta x_{n+1}-2\delta x_n+\delta x_{n-1})\approx \partial^2 \delta x/\partial n^2$ as in the Rouse model~\cite{kaiser2015does}. We have done this, and the results can well describe the long-wavelength behaviors, but when $\mu k\tau\ll 1$, the high-frequency terms become more and more important, and such a long-wavelength approximation breaks down. Therefore, we introduce a standard discrete Fourier transform to the equations of motion for an $N$-particle chain, as given in Eq.~\eqref{eq:EOM_PBCs}:%
\begin{equation*}
    \tilde{x}_q=\frac{1}{N}\sum_{n=0}^{N-1}\delta x_n\mathrm{e}^{-iqn},
\end{equation*}
where $q=2\pi m/N$ and the index $m=0,1,\cdots,N-1$. The inverse transform is given by $\delta x_n=\sum_q\tilde{x}_q\mathrm{e}^{iqn}$.
By substituting the inverse transforms into the above Eq.~\eqref{eq:EOM_PBCs} and matching terms for each mode $q$, we can then obtain
\begin{equation}\label{eq:EOM_q_appendix}
    \dot{\tilde{x}}_q = -\mu k_q\tilde{x}_q+ \tilde{v}_q,
\end{equation}
where $k_q=2k(1-\cos q)$ and $\tilde{v}_q=(1/N)\sum_{n=0}^{N-1}v_n\mathrm{e}^{-iqn}$ is the Fourier amplitude of $v_n$. The mean is simply $\langle \tilde{v}_q\rangle = 0$, and the time correlation is
\begin{eqnarray}
    \langle \tilde{v}^{\phantom{\ast}}_q(t)\tilde{v}^\ast_{q'}(t')\rangle &=&\frac{1}{N^2}\sum_{n=0}^{N-1}\sum_{n'=0}^{N-1}\langle v_n(t) v_{n'}(t')\rangle\mathrm{e}^{-iqn+iq'n'} \nonumber\\
    &=& \delta_{qq'}\frac{D}{N\tau}\mathrm{e}^{-|t-t'|/\tau},
\end{eqnarray}
where we have used the mean and time correlation of $v_n$, along with the orthogonality relation $\sum_{n=0}^{N-1}\mathrm{e}^{i(q-q')n}=N\delta_{qq'}$.
Equation~\eqref{eq:EOM_q_appendix} can be solved by using the exponential ansatz $\delta\tilde{x}_q = \mathcal{R}_q(t)\mathrm{e}^{-\mu k_q t}$. When $t\gg 1/\mu k_q$, we have
\begin{equation*}
    \tilde{x}_q(t)=\int_0^t\mathrm{d} t'~ \mathrm{e}^{-\mu k_q(t-t')}\tilde{v}_q(t').
\end{equation*}
Thus, for the zeroth mode $\tilde{x}_0=(1/N)\sum_n \delta x_n$ (center of mass), {which is special because $k_0 = 0$}, we obtain $\langle \tilde{x}_0\rangle=0$ and $\langle| \tilde{x}_0(t)|^2\rangle = 2D t/N$. For other modes with $q\neq 0$, $\langle \tilde{x}_q\rangle = 0$ and
\begin{equation}
    \langle \tilde{x}^{\phantom{\ast}}_q(t)\tilde{x}^\ast_{q'}(t)\rangle =\delta_{qq'}\frac{D}{N\mu k_q(1+\mu k_q\tau)}.
\end{equation}
Ref.~\cite{henkes2020dense} shows that active Brownian particles (ABPs) exhibit a similar relationship, as their active velocities share the same Gaussian colored noise temporal correlations as those of AOUPs.
Then we can calculate the stretched lengths of the springs by $\delta\ell_n = x_{n+1} - x_n-\ell_0=\delta x_{n+1}-\delta x_n=\sum_{q\neq 0}\tilde{x}_q\mathrm{e}^{iqn}(\mathrm{e}^{iq}-1)$, where the mean is simply $\langle\delta\ell_n\rangle = 0$, and the variance is
\begin{eqnarray}
    \langle \delta\ell_n^2\rangle
    &=&\sum_{q\neq 0 }\sum_{q'\neq 0}\langle \tilde{x}^{\phantom{\ast}}_q\tilde{x}^\ast_{q'}\rangle \mathrm{e}^{i(q-q')n}(\mathrm{e}^{iq}-1)(\mathrm{e}^{-iq'}-1)\nonumber\\
    &=&\frac{D}{N\mu k}\sum_{q\neq 0}\frac{1}{1+2\mu k\tau (1-\cos q)}.\nonumber
\end{eqnarray}
In the limit $N\gg 1$, we can replace the sum by an integral
\begin{equation*}
    \sum_{q\neq 0}\frac{1}{1+2\mu k\tau (1-\cos q)} \approx\frac{N}{2\pi}\int_0^{2\pi}\frac{\mathrm{d} q}{1+2\mu k\tau (1-\cos q)},
\end{equation*}
which equals $N/\sqrt{1+4\mu k\tau}$.
Then, we obtain the variance of $\delta\ell_n$ given in the main text:
\begin{equation}
    \langle \delta\ell_n^2\rangle =\frac{D}{\mu k\sqrt{1+4\mu k\tau}}.
\end{equation}
Thus, in the steady state, the marginal distribution of $\delta\ell_n$ follows a Gaussian distribution $P_s(\delta\ell_n)=\mathcal{Z}\exp(-\delta\ell_n^2/2\langle\delta\ell_n^2\rangle)$ since each  $\delta \ell_n$ is the sum of Gaussian processes, and each $\delta x_n$ is similarly a Gaussian process.
In the small noise limit, the system rapidly relaxes from $P_s(\delta\ell_n)$ to a quasi-stationary state described by $P_\mathrm{QS}(\delta\ell_n,t )\sim P_s(\delta\ell_n)\,\mathrm{e}^{-t/\tau_\mathrm{esc}}$, indicating that the escape process is Poissonian with a rate of $1/\tau_\mathrm{esc}$~\cite{woillez2020active}. 
Therefore, the mean escape time follows an effective ``Arrhenius law'' where the rate of rupture is proportional to the probability density at the breaking stretch $\delta \ell_b$, $\tau_\mathrm{esc}^{-1}\sim P_s(\delta\ell_b)$~\cite{woillez2020active}.
We then can find the mean escape time is given by
\begin{equation*}
\tau_\textrm{esc}\sim\exp\left(\frac{\mu k\delta\ell_b^2\sqrt{1+4\mu k\tau}}{2D}\right).
\end{equation*}
{We propose the result
\begin{equation}
\tau_\textrm{esc}=\frac{2\pi\eta}{\omega_0^2\tau}\exp\left(\frac{\mu k\delta\ell_b^2\sqrt{1+4\mu k\tau}}{2D}\right).
\end{equation}
We originally derived this prefactor as $2\pi\eta/\sqrt{U''(0)|U''(\delta\ell_b)|}$, which is the prefactor appropriate to an overdamped thermal escape~\cite{hanggi1990reaction}. %
However, its application to this case, where the potential is a cusp, is somewhat \textit{ad hoc}---we view this as a reasonable first approximation, and it has a good fit to the simulated rates.
While the prefactor may be slightly different under different parameter regimes, the leading behavior for small effective temperatures is dominated by the exponential term~\cite{woillez2020active}---so the prefactor is essentially only serving to set the units of the problem, and its magnitude can be neglected.}

\subsection{When is the two-cell result the \textit{N}-cell result?---an effective equipartition theorem}\label{app:equipartition}

For the thermal equilibrium case where $\tau=0$, the variances of stretched lengths for both a cell pair $\langle\delta\ell^2\rangle$ and a chain $\langle\delta\ell_n^2\rangle$ are identical, given by $\langle\delta\ell^2\rangle_\mathrm{th}=D/\mu k$.
This result can be easily verified using the equipartition theorem. For a long chain, the energy of the system is expressed as $\Phi=(1/2)k\sum_n\delta\ell_n^2$. Due to translational symmetry in the limit $N\gg 1$, we have
\begin{equation*}
    \langle\Phi\rangle = \frac{N}{2}k\langle\delta\ell_n^2\rangle = \frac{N}{2}k_B T,
\end{equation*}
where $k_B T=D/\mu$ from the Einstein relation, thus we obtain $\langle\delta\ell_n^2\rangle = D/\mu k$. For a particle pair, $\Phi = (1/2)k(x_2-x_1-\ell_0)^2 = (1/2)k\delta\ell^2$, we then get the same result by applying the equipartition theorem.

However, we note that the results for both the pair and the chain also converge to the same non-equilibrium value when $\mu k\tau \ll 1$ because $\sqrt{1+4\mu k\tau}\to 1+2\mu k\tau$ for small $\mu k \tau$. Why should an active system with a finite persistence time $\tau$ (nonequilibrium) still show that $\langle \delta \ell^2\rangle_\textrm{pair} = \langle \delta \ell_n^2\rangle_\textrm{chain}$, as if some equipartition theorem applied? In Ref.~\cite{fodor2016far}, Fodor \textit{et al.} found that {for a system of interacting AOUPs}, at small but finite persistence time $\tau$, there exists an \emph{effective equilibrium} regime---the particle dynamics still respects detailed balance and time-reversal symmetry---the entropy production rate vanishes. Rather than defining an effective temperature as in Appendix~\ref{app:two_particle}, the authors stated that the system could be equivalently described using an effective potential, while defining $k_B T =D/\mu$ ~\cite{fodor2016far,martin2021statistical}:
\begin{equation}
   \tilde{\Phi} = \Phi + \tau\left[\frac{\mu}{2}(\nabla_n \Phi)^2 - D\nabla^2_n \Phi\right] +\mathcal{O}(\tau^2).
\end{equation}
{Note that we have restored the mobility $\mu$ which was set to one in Refs.~\cite{fodor2016far,martin2021statistical},} and $\nabla_n \equiv \partial/\partial{x_n}$ denotes the derivative with respect to $x_n$, not $\delta\ell_n$. The Einstein summation convention is applied here, i.e., $(\nabla_n\Phi)^2 = \sum_n(\partial_{x_n}\Phi)^2$ and $\nabla_n^2\Phi =\sum_n\partial_{x_n}^2\Phi$. For a particle pair, we have $(\nabla_n\Phi)^2 = 2k^2(x_2-x_1-\ell_0)^2=2k^2\delta\ell^2$, and $\nabla_n^2\Phi = 2k$, which is a constant and can therefore be disregarded. Applying equipartition on this effective potential,
\begin{equation*}
    \langle \tilde{\Phi}\rangle = \frac{1}{2}k (1+2\mu k\tau) \langle\delta\ell^2\rangle=\frac{1}{2}k_B T,
\end{equation*}
the variance derived in Appendix~\ref{app:two_particle} follows immediately. For a long chain, it yields $(\nabla_n \Phi)^2=k^2\sum_n(x_{n+1}-2x_n+x_{n-1})^2=k^2\sum_n(\delta\ell_{n+1}-\delta\ell_n)^2$, and $\nabla_n^2 \Phi =2Nk$, which is also a constant. Therefore, we have
\begin{equation*}
     \frac{N}{2} k\left[\langle\delta\ell_n^2\rangle + {2\mu k \tau}\left(\langle\delta\ell_{n}^2\rangle - \langle\delta\ell_{n}\delta\ell_{n+1}\rangle\right)\right]=\frac{N}{2}k_B T,
\end{equation*}
where we have used the translational symmetry of the system. We can directly calculate $\langle \delta\ell_n\delta\ell_{n+1}\rangle$, the correlation between adjacent stretched lengths as
\begin{eqnarray}
    &&\sum_{q\neq 0 }\sum_{q'\neq 0}\langle \tilde{x}^{\phantom{\ast}}_q\tilde{x}^\ast_{q'}\rangle \mathrm{e}^{iqn-iq'(n+1)}(\mathrm{e}^{iq}-1)(\mathrm{e}^{-iq'}-1)\nonumber\\
    &=&\frac{D}{N\mu k}\sum_{q\neq 0}\frac{\cos q}{1+2\mu k\tau (1-\cos q)}\nonumber\\
    &\approx& \frac{D}{2\pi\mu k}\int_0^{2\pi}\mathrm{d} q~\frac{\cos q}{1+2\mu k\tau(1-\cos q)}\nonumber\\
    &=&\langle\delta\ell_n^2\rangle\frac{1+2\mu k\tau -\sqrt{1+4\mu k\tau}}{2\mu k\tau}.\nonumber
\end{eqnarray}
Since $\sqrt{1+2x}=1+x-x^2/2+\mathcal{O}(x^3)$, we have $\langle\delta\ell_n\delta\ell_{n+1}\rangle \approx \mu k\tau\langle\delta\ell_n^2\rangle\ll\langle\delta\ell_n^2\rangle$ when $\mu k\tau\ll 1$. Thus, the above equation for a chain yields the two-particle result $\langle\delta\ell_n^2\rangle \approx D/\mu k(1+2\mu k\tau)$ when $\mu k\tau\ll 1$.

\section{Solving the master equations for division and fracture with generating functions}\label{app:gen_func}

\subsection{All particles can divide}\label{app:gen_func_1}
To solve the master equation~\eqref{eq:master_all} in the main text, we first define a generating function as
\begin{equation}
    G(z,t)\coloneq\sum_{n\geqslant 1} p_n(t) z^n.
\end{equation}
We then multiply the master equation by $z^n$ and sum over all $n$, so the left-hand side becomes $\partial G(z,t)/\partial t$. On the right-hand side, the first term is 
\begin{eqnarray}
&&k_b\sum_{n\geqslant 1}z^n\cdot\sum_{m>n} p_m \nonumber\\
&=& k_b\sum_{m>1}p_m\cdot\sum_{n=1}^{m-1}z^n = k_b\sum_{m>1}p_m\frac{z-z^{m}}{1-z}\nonumber\\
&=&k_b\frac{z}{1-z}\sum_{m>1}p_m-k_b\frac{1}{1-z}\sum_{m>1}p_mz^{m}\nonumber\\
&=&\frac{k_bz}{1-z}(1-p_1(t))-\frac{k_b}{1-z}(G(z,t)-p_1(t)z)\nonumber\\
&=&\frac{k_b}{1-z}\left(z-G(z,t)\right),\nonumber
\end{eqnarray}
 where we interchange the order of summation in the first step, and we have used the normalization condition $\sum_m p_m=1$ in the penultimate step. The second term is
\begin{equation*}
    -k_b\sum_{n\geqslant 1}z^n(n-1)p_n=-k_b\left(z\frac{\partial G}{\partial z}-G\right),
\end{equation*}
and the third term is
\begin{equation*}
    k_d\sum_{n\geqslant 1}z^n(n-1)p_{n-1}=k_d z^2\frac{\partial}{\partial z}\sum_{n\geqslant 1}p_{n-1}z^{n-1}=k_dz^2\frac{\partial G}{\partial z},
\end{equation*}
where we have used the condition $p_n(t)\equiv 0$ when $n=0$. The last term is
\begin{equation*}
    -k_d\sum_{n\geqslant 1}z^nnp_n=-k_d z\frac{\partial G}{\partial z}.
\end{equation*}
Hence, we obtain a partial differential equation (PDE) for the generating function:
\begin{equation}
    \frac{\partial G}{\partial t}=(k_dz -k_b-k_d)z\frac{\partial G}{\partial z}-k_b\frac{z}{1-z}G+k_b\frac{z}{1-z},
\end{equation}
where the steady-state solution is 
\begin{equation}\label{eq:G_steadystate_intermediate}
    G(z)=\frac{k_b-(1-z)C}{k_b+k_d-k_dz}, 
\end{equation}
where $C$ is a constant we have to set. Expanding $G(z)$ in powers of $z$, we get
\begin{equation*}
G(z)=\frac{k_b-C}{k_b+k_d}+\frac{k_b(C+k_d)}{(k_b+k_d)^2}z+\mathcal{O}(z^2).
\end{equation*}
By construction, our function $G(z)$ has no zeroth-order term, so the constant must be $C=k_b$. %
We could also find this constant by explicitly solving the original master equation for $p_1$: $\mathrm{d}p_1/\mathrm{d}t=\sum_{m>1}k_bp_m-k_dp_1=k_b(1-p_1)-k_dp_1$. In the steady state, setting $\mathrm{d}p_1/\mathrm{d}t=0$ gives $p_1=k_b/(k_b+k_d)$. This result holds for all three growth models, as the master equation for $n=1$ remains the same in each case.
Matching the first-order coefficient to $p_1$ gives the same value of $C=k_b$. Using this value of $C$ in Eq.~\eqref{eq:G_steadystate_intermediate}, we find the steady-state generating function is given by
\begin{equation}
    G(z)=\frac{k_b z}{k_b+k_d-k_d z}=\sum_{n\geqslant 1}\frac{k_b}{k_d}\left(\frac{k_d}{k_b+k_d}\right)^n z^n.
\end{equation}
We can read the value of $p_n$ directly from this formal series, finding Eq.~\eqref{eq:ss_solution_all} in the main text:
\begin{equation}\label{eq:app_pn_all}
    p_n = \frac{k_b}{k_d}\left(\frac{k_d}{k_b+k_d}\right)^n = \frac{\gamma}{(1+\gamma)^n},
\end{equation}
where we define the ratio $\gamma\coloneq k_b/k_d$. When $\gamma\to0$, $p_n\approx\gamma(1-n\gamma)$; when $\gamma\gg 1$, $p_n\approx1/\gamma^{n-1}$ decreases exponentially with $n$.
The first two moments are
\begin{subequations}
\begin{eqnarray}
    \langle n \rangle &=& \sum_{n} p_n n = \frac{1+\gamma}{\gamma},\\
    \langle n^2\rangle&=&\sum_{n} p_n n^2=\frac{(1+\gamma)(2+\gamma)}{\gamma^2},
\end{eqnarray}
\end{subequations}
and the variance is
\begin{equation}
    \langle n^2\rangle-\langle n\rangle^2 = \frac{1+\gamma}{\gamma^2}=\frac{\langle n\rangle}{\gamma}.
\end{equation}
We can also calculate the coefficient of variation ($\mathrm{CV}$) as
\begin{equation}
    \mathrm{CV}\coloneq\frac{\sigma_n}{\mathbb{E}[n]}=\frac{ \sqrt{\langle n^2\rangle-\langle n\rangle^2}}{\langle n\rangle}=\frac{1}{\sqrt{1+\gamma}},
\end{equation}
where $\sigma_n$ represents the standard deviation of $n$, and $\mathbb{E}[n]$ denotes the expected value of $n$.

\begin{figure*}[t]
\includegraphics[width=0.85\textwidth]{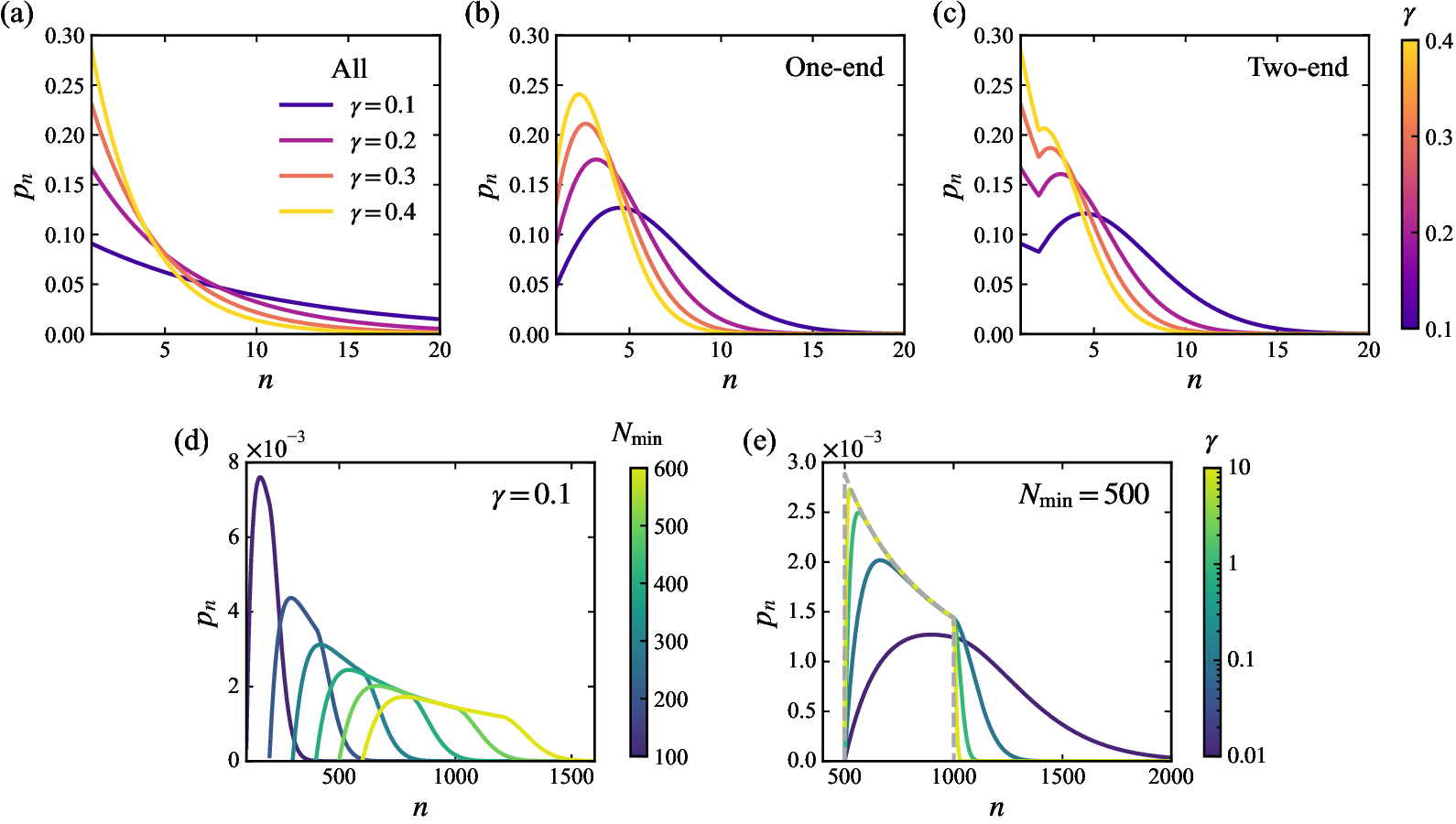}
\caption{How the steady-state distributions $p_n$ vary with $\gamma$ and $N_{\min}$ across different growth models.
(a) Steady-state distribution $p_n$ for various values of $\gamma$ in the all-cell growth model [Eq.~\eqref{eq:app_pn_all}].
(b) Steady-state distribution $p_n$ for different $\gamma$ in the one-end growth model, where we have doubled the division rate [$\gamma\to\gamma'=\gamma/2$ in Eq.~\eqref{eq:app_pn_1}].
(c) Steady-state distribution $p_n$ for different $\gamma$ in the two-end growth model from Eq.~\eqref{eq:app_pn_2}.
(d) Steady-state distribution $p_n$ obtained from the matrix equation $\bm{K}\mathbf{p}=\mathbf{0}$ incorporating different values of minimum cluster size $N_\mathrm{min}$ with $\gamma=0.1$.
(e) Steady-state distribution $p_n$ obtained from the matrix equation method for different values of $\gamma$ with $N_\mathrm{min}=500$. The gray dashed line represents the corresponding log-uniform distribution $1/(n\ln2)$ within the range $[N_\mathrm{min}, 2N_\mathrm{min}]$.
}
\label{fig:sm_distrib_all_models}
\end{figure*}

\subsection{One-end growth}\label{app:gen_func_2}
To solve Eq.~\eqref{eq:master_one} in the main text, we can still apply the generating function method. The first two terms on the right-hand side {are the same as in the all-cell growth case, but the last two terms after multiplication by $z^n$ and summing over $n$ become}
\begin{equation*}
    k_d\sum_{n\geqslant 1}z^n p_{n-1}=k_dz\sum_{n\geqslant 1}z^{n-1} p_{n-1}=k_dzG,
\end{equation*}
and
\begin{equation*}
    -k_d\sum_{n\geqslant 1}z^n p_n=-k_d G.
\end{equation*}
Hence, the PDE for $G(z,t)$ for the one-end-growth model is  given by
\begin{equation*}
    \frac{\partial G}{\partial t}=-k_b z \frac{\partial G}{\partial z} - \left(k_d(1-z)+k_b\frac{z}{1-z}\right)G+k_b\frac{z}{1-z}.
\end{equation*}
Then the steady-state solution is
\begin{equation*}
    G(z)=\frac{(1-z)}{z^{1/\gamma}}\mathrm{e}^{z/\gamma}\left(\int_1^z\frac{x^{1/\gamma}}{(1-x)^2}\mathrm{e}^{-x/\gamma}\mathrm{d}x+C\right).
\end{equation*}
Similarly, we can determine the constant $C$ by matching the conditions $p_0=0$ and $p_1=k_b/(k_b+k_d)$:
\begin{equation*}
    C=\int_0^1\frac{x^{1/\gamma}}{(1-x)^2}\mathrm{e}^{-x/\gamma}\mathrm{d}x,
\end{equation*}
so the generating function is given by
\begin{equation}
    G(z)=\frac{(1-z)}{z^{1/\gamma}}\mathrm{e}^{z/\gamma}\int_0^z\frac{x^{1/\gamma}}{(1-x)^2}\mathrm{e}^{-x/\gamma}\mathrm{d}x.
\end{equation}
Expanding $G(z)$ around $z=0$, we obtain Eq.~\eqref{eq:ss_solution_one} in the main text:
\begin{equation}\label{eq:app_pn_1}
    p_n=\frac{n\gamma}{\prod_{m=1}^n(1+m\gamma)}=\frac{n\gamma^{1-n}}{\bm{(}(1+\gamma)/\gamma\bm{)}_n},
\end{equation}
where $(x)_n=x(x+1)(x+2)\cdots(x+n-1)=\sum_{k=0}^{n-1}(x+k)$ denotes the Pochhammer symbol. %
The distribution $p_n$ for one-end growth is shown in Fig.~\ref{fig:sm_distrib_all_models}(b), where a distinct peak (or typical scale of $n$) is observed.
When $\gamma\gg 1$, $p_n\approx \gamma^{1-n}/(n-1)!$ deceases faster than the all-cell growth results. Since $n=1$ becomes the most probable outcome when $\gamma$ is large, statistical quantities such as the mean and variance should converge to those of the all-cell growth model.
For small $\gamma$, to determine the location of the peak in $p_n$, we compute the ratio:
\begin{equation*}
    \frac{p_{n+1}}{p_n}=\frac{n+1}{n}\frac{1}{1+(n+1)\gamma}.
\end{equation*}
By setting $p_{n+1}/p_n=1$, we arrive at the equation $n(n+1)=1/\gamma$, which implies the peak $n^\ast\approx 1/\sqrt{\gamma}$, and $n^\ast\gg 1$ when $\gamma\ll 1$.
The log-probability function is given by
$$\ln p_n=\ln(n\gamma)-\sum_{m=1}^n\ln(1+m\gamma).$$
For large $n$ and small $\gamma$, the sum can be approximated by
$\int_0^n\ln(1+m\gamma) \,\mathrm{d}m\approx\int_0^n m \gamma \,\mathrm{d}m= {\gamma} n^2/2$.
Thus, when $\gamma\ll 1$, the distribution near the peak $n=n^\ast$ can be approximated as
\begin{equation}
    p_n\approx f(n;\sigma)=n\gamma \mathrm{e}^{-\gamma n^2/2},
\end{equation}
where $f(n;\sigma)$ is a Rayleigh distribution with mode $\sigma=1/\sqrt{\gamma}$.
Therefore, when $\gamma\ll 1$, we can use the saddle-point approximation to compute the average of any $X(n)$:
\begin{eqnarray*}
    \langle X(n)\rangle&=&\sum_n p_n X(n)\approx \int_0^\infty p_n X(n)\mathrm{d} n\\
    &\approx&\int_0^\infty f(n;\sigma) X(n)\mathrm{d} n.
\end{eqnarray*}
Thus, the mean and variance of $n$ are directly obtained from the Rayleigh distribution:
\begin{equation}
    \langle n\rangle=\sigma\sqrt{\frac{\pi}{2}},~~~\langle n^2\rangle -\langle n\rangle^2 = \frac{4-\pi}{2}\sigma^2,
\end{equation}
and the coefficient of variation is
\begin{equation}
    \mathrm{CV}=\frac{\sqrt{\langle n^2\rangle -\langle n\rangle^2}}{\langle n\rangle} = \sqrt{\frac{4-\pi}{\pi}}.
\end{equation}

\subsection{Two-end growth}\label{app:gen_func_3}
For the two-end growth scenario, the master equation is described by Eq.~\eqref{eq:master_all} for $n=1$ and $n=2$, and by Eq.~\eqref{eq:master_one} with a doubled division rate for $n>2$: \begin{eqnarray}
\frac{\mathrm{d} p_n}{\mathrm{d} t} = 
\begin{cases}
    \mathcal{F}(k_b)+0-k_dp_n, & n=1,\nonumber\\
    \mathcal{F}(k_b)+k_dp_{n-1}-2k_dp_n, & n=2,\nonumber\\
    \mathcal{F}(k_b)+2k_dp_{n-1}-2k_dp_n, & n>2,\nonumber
\end{cases}
\end{eqnarray}
where $\mathcal{F}(k_b)={k_b}\sum_{m>n}p_m-k_b(n-1)p_n$ represents the fragmentation process which remains unchanged. Using the generating function method, {multiplying the master equation by $z^n$ and summing over $n$, the generating function corresponding to $\mathcal{F}(k_b)$ are just the ones we derived before.} However, the last two terms in the master equation become
\begin{equation}
0+k_dp_1z^2+2k_d\sum_{n>2}p_{n-1}z^n=-k_dp_1 z^2 + 2k_d z G,\nonumber
    \end{equation}
    and
    \begin{equation}
    -k_dp_1 z - 2k_d\sum_{n\geqslant 2}p_nz^n=k_dp_1 z - 2k_d G.\nonumber
\end{equation}
In the steady state, we have $p_1=k_b/(k_b+k_d)$ and $\partial G/\partial t=0$, so we obtain the equation:
\begin{eqnarray}
    k_b z \frac{\partial G}{\partial z}&=&-\left(2k_d(1-z)+k_b\frac{z}{1-z}\right)G+k_b\frac{z}{1-z}\nonumber\\
    &&+\frac{k_bk_d}{k_b+k_d}z(1-z).
\end{eqnarray}
The solution is given by
\begin{equation*}
    G(z)=\frac{(1-z)}{z^{2/\gamma}}\mathrm{e}^{2z/\gamma}\int_0^z\frac{x^{2/\gamma}(x^2-2x+2+\gamma)}{(1+\gamma)(1-x)^2}\mathrm{e}^{-2x/\gamma}\mathrm{d}x,
\end{equation*}
where the integration constant has been set as
\begin{equation*}
    C = \int_0^1\frac{x^{2/\gamma}(x^2-2x+2+\gamma)}{(1+\gamma)(1-x)^2}\mathrm{e}^{-2x/\gamma}\mathrm{d}x,
\end{equation*}
to match $p_0=0$, and $p_1=k_b/(k_b+k_d)$. By expanding $G(z)$ around $z=0$, we derive the exact analytical solution for two-end growth case:
\begin{equation}\label{eq:app_pn_2}
    p_n = 
    \begin{cases}
        \gamma/(1+\gamma), & n=1,\\
        \dfrac{2+\gamma}{2(1+\gamma)}\dfrac{n(\gamma/2)^{1-n}}{\bm{(}(2+\gamma)/\gamma\bm{)}_n}, & n\geqslant2.
    \end{cases}
\end{equation}
Observe that the results are quite similar to the one-end growth case, but with a doubled division rate $2k_d$ [equivalent to replacing $\gamma$ with $\gamma/2$ in Eq.~\eqref{eq:app_pn_1}]. When $\gamma \to 0$, meaning only growth without fragmentation, the two scenarios converge. In contrast, when $\gamma \gg 1$, the chain is most likely to be a monomer ($p_1 \approx 1$), and $p_n$ is reduced by half compared to the one-end growth scenario with $2k_d$ for $n\geqslant 2$.

\section{Generalizations to the model with a matrix equation method}\label{app:matrix}

A general master equation with a minimum cluster size $N_\textrm{min}$ and different growth mechanisms can be written as
\begin{widetext}
\begin{equation}\label{eq:generalmaster}
   \frac{\mathrm{d} p_n}{\mathrm{d} t}=k_b\sum_{m\geqslant n+N_{\min}}p_m-k_b\max(n-2N_{\min}+1, 0)p_n+g(n-1)p_{n-1}-g(n)p_n,~~~(n\geqslant N_\textrm{min})
\end{equation}
\end{widetext}
where we have incorporated in the first term that given a minimum cluster size $N_\textrm{min}$, a cluster of size $m$ can only break into one of size $n$ if $m\geqslant n+N_\textrm{min}$. In the second term, we require that both daughter clusters must satisfy $i,j\geqslant N_\textrm{min}$, i.e., this term is
\begin{equation}
p_n\sum_{\substack{i+j=n\\i,j\geqslant N_{\textrm{min}}}}k_b=k_b\max(n-2N_\textrm{min}+1, 0) p_n,
\end{equation}
where $\max$ acts as an activation function indicating that the chain can only break once it has grown to at least $2N_\textrm{min}$.
In the third and fourth terms of our general master equation [Eq.~\eqref{eq:generalmaster}], $g(n)$ describes how the total growth rate scales with the length of the chain:
\begin{equation}
    g(n)\coloneq
    \begin{cases}
        k_d n, & \text{all-cell,}\\
        k_d, & \text{one-end,}\\
        k_d[1+\Theta(n-2)], & \text{two-end,}
    \end{cases}
\end{equation}
where $\Theta(x)$ is the Heaviside step function, {with the convention $\Theta(0)=1$ adopted throughout this paper.}
Then, the master equation can be organized as
\begin{equation}\label{eq:master_general}
    \frac{\mathrm{d} p_n}{\mathrm{d} t}=k_b\sum_{m\geqslant n+N_{\min}}p_m  + g(n-1)p_{n-1} - \lambda(n)p_n,
\end{equation}
where $\lambda(n)\coloneq k_b\max(n-2N_{\min}+1, 0)+g(n)$ is the total transition rate from $n$-mer state to other states.
As a continuous-time Markov chain, Eq.~\eqref{eq:master_general} can be written in the matrix form as $\dot{\mathbf{p}} = \bm{K}\mathbf{p}$, where $\bm{K}$ is the generator on the state space $\{N_\textrm{min}, N_{\textrm{min}}+1, N_{\textrm{min}}+2,\cdots\}$. For simplicity, we decompose $\bm{K}=\bm{A}+\bm{B}$, where $\bm{A}$ is a smaller upper triangular matrix:
\begin{equation*}
\bm{A}\mathbf{p}=
    \begin{pmatrix}
        0&0&0&\cdots&k_b&k_b&k_b&\cdots \\
        0&0&0&\cdots&0&k_b&k_b&\cdots\\
        0&0&0&\cdots&0&0&k_b&\cdots\\
        \vdots&\vdots&\vdots&\ddots&&&&\ddots\\
        &\\
        &\\
        &\\
        0&&&&&&&\\
    \end{pmatrix}
    \begin{pmatrix}
        p_{N_{\min}\phantom{+0}} \\
        p_{N_{\min}+1}\\
        p_{N_{\min}+2}\\
        \vdots\\
        p_{2N_{\min}\phantom{+0}}\\
        p_{2N_{\min}+1}\\
        p_{2N_{\min}+2}\\
        \vdots
    \end{pmatrix},
\end{equation*}
and $\bm{B}$ is a lower bidiagonal matrix:
\begin{equation*}
\bm{B}=
    \begin{pmatrix}
        -\lambda(N_{\min}) & 0 & 0 & \cdots \\
        \phantom{-}g(N_{\min})&-\lambda(N_{\min}+1) & 0 & \cdots\\
        0 & \phantom{-}g(N_{\min} + 1) & -\lambda(N_{\min}+2) & \cdots\\
        0 & 0 & \phantom{-}g(N_{\min}+ 2) & \cdots\\
        \vdots & \vdots & \vdots & \ddots
    \end{pmatrix},
\end{equation*}
i.e., the main diagonal $B_{nn}=-\lambda(n)$ and subdiagonal $B_{n+1, n}=g(n)$. Note that the transition-rate matrix $\bm{K}$ satisfies the property $K_{nn}=-\sum_{m\neq n}K_{mn}$, namely, the columns of the matrix sum to zero.
In the steady state, we obtain a homogeneous equation $\bm{K}\mathbf{p}=\mathbf{0}$.
We can then find the null space of $\bm{K}$ and normalize it to obtain the steady-state distribution $\mathbf{p}$, ensuring that $\sum_{n\geqslant N_\textrm{min}}p_n=1$. 
{Note that to solve this equation numerically, we must truncate to a maximum possible cluster size $N_\mathrm{max}\gg 2N_\mathrm{min}$, where $p_n$ is negligible for $n>N_\mathrm{max}$.}
$N_\textrm{max}$ is set to $N_\mathrm{min}+5000$ by default in our calculations. 
It is also possible to incorporate the normalization condition into the matrix $\bm{K}$ by adding an extra row $(1,1,1,\cdots)$ at the top, forming a new matrix $\bm{K}'$.
Then the matrix equation becomes
$\bm{K}'\mathbf{p}=\mathbf{c}$, where $\mathbf{c}=(1,0,0,\cdots)^\top$. Note that $\bm{K}'$ is not a square matrix after adding the normalization row. However, by multiplying the matrix equation by $\bm{K}'\vphantom{\bm{K}}^{\top}$ from the left, the product $\bm{K}'\vphantom{\bm{K}}^{\top}\bm{K}'$ becomes square again, allowing us to compute its inverse. The final solution is given by $\mathbf{p} = \left(\bm{K}'\vphantom{\bm{K}}^{\top}\bm{K}'\right)^{-1}\bm{K}'\vphantom{\bm{K}}^{\top}\mathbf{c}$.
When setting $N_{\min}=1$, the solutions $\mathbf{p}$ match those derived in Appendix~\ref{app:gen_func} using the generating function method (see Fig.~\ref{fig:suppl_matrix}).

\begin{figure}
\hspace{-0.5cm}
\includegraphics[width=0.48\textwidth]{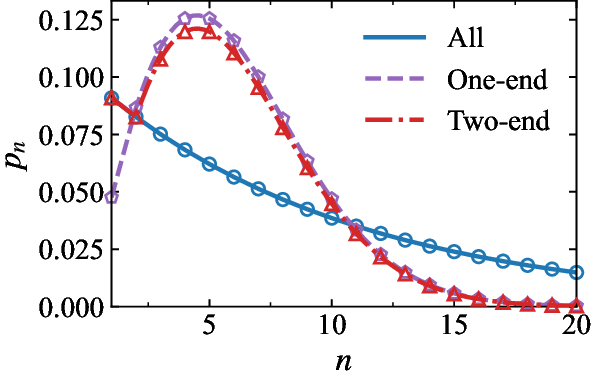}
\caption{Comparison of the analytical solution obtained using the generating function method with numerical results from the matrix equation method. Lines represent the same analytical solutions for $\gamma=0.1$ as shown in Fig.~\ref{fig:statistics}(a) in the main text ($\gamma'=\gamma/2=0.05$ for the one-end growth model). Empty symbols denote exact numerical results from the matrix equation $\bm{K}\mathbf{p}=\mathbf{0}$.}
\label{fig:suppl_matrix}
\end{figure}

\subsection{Basal rates}\label{app:basal}
So far we have assumed that both the end-growth and minimum cluster size models are strictly enforced, meaning that break and division rates are either at their maximum values or completely suppressed. However, in real cell clusters, it is not clear whether breakage would be completely absent, even if its rate is larger near the cluster center. Similarly, cells within a cluster may still divide despite contact inhibition of proliferation. Therefore, it is reasonable to incorporate basal rates into our model---changing the model so that every cell has a basal rate of division and every junction has a basal rate of fracture.
This is straightforward with the matrix equation method. 

We first consider a scenario where, in addition to the regulated break rate $k_b$ (restricted by the minimum cluster size), there exists a basal break rate $k_b^0$ that applies uniformly to all links. In our matrix equation formulation, we solve $\bm{K}\mathbf{p}=\mathbf{0}$, where now $\mathbf{p}=(p_1,p_2,p_3,\cdots)^\top$, rather than $(p_{N_\mathrm{min}}, p_{N_\mathrm{min}+1}, p_{N_\mathrm{min}+2}, \cdots)^\top$.
We continue to decompose the transition-rate matrix as $\bm{K}=\bm{A}+\bm{B}$, where $\bm{A}$ remains an upper triangular matrix, but with modified elements $A_{mn} = \Theta(n-m-1) k_b^0 + \Theta(m-N_\mathrm{min})\Theta(n-m-N_\mathrm{min})k_b$. Note that the second term corresponds exactly to the original matrix $\bm{A}$.
$\bm{B}$ remains the bidiagonal matrix but is updated with the total transition rate $\lambda(n)=k_b^0(n-1)+k_b\max(n-2N_\mathrm{min}+1,0)+g(n)$.
Figure~\ref{fig:suppl_basal}(a) illustrates how the steady-state distributions change, as in Fig.~\ref{fig:sm_distrib_all_models}(e), when a small basal break rate $k_b^0=10^{-3}k_b$ is applied to all links, while keep the definition $\gamma\coloneq k_b/k_d$.
For small $k_b^0$ (low $\gamma$), the distribution remains nearly unchanged, but as $k_b^0$ increases with increasing $\gamma$ (near-deterministic), it effectively relaxes the minimum cluster size constraint, shifting the distribution toward an exponential decay.
Figure~\ref{fig:suppl_basal}(b) further demonstrates this effect for a near-deterministic break case ($\gamma=10$), where the distribution gradually leaks into $n<N_\mathrm{min}$. {We note that the basal break rate does not have to be very large in order to qualitatively change our results---even if the basal break rate is a factor of 100 lower than $k_b$, it can overwhelm the ``controlled'' break rate that respects $N_\textrm{min}$, leading to a system dominated by small clusters, as when all links break at the same rate. This may argue that the limit of fast rupture $\gamma \gg 1$ is not a biologically robust way to control size, since it requires a very small basal break rate.}

\begin{figure}
\includegraphics[width=0.48\textwidth]{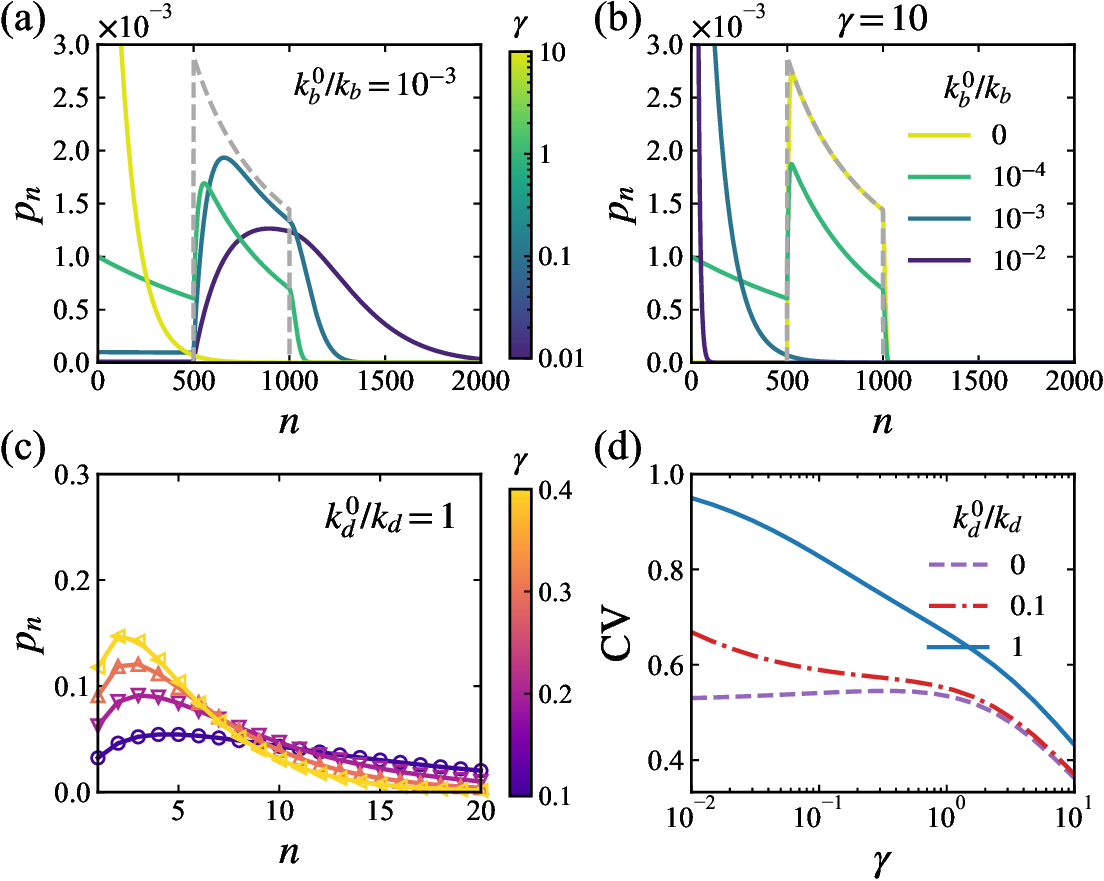}
\caption{Effect of basal break and division rates on the steady-state distributions $p_n$.
(a) Cluster sizes when $N_\textrm{min} = 500$ and $\gamma$ is varied, but with a basal break rate $k_b^0=10^{-3}k_b$, i.e., the basal break rate varies when $\gamma$ changes; this figure should be compared to  Fig.~\ref{fig:sm_distrib_all_models}(e).
(b) The near-deterministic case $\gamma=10$ with varying basal break rates $k_b^0$.
Minimum cluster size is set to $N_\mathrm{min}=500$; gray dashed lines are the corresponding log-uniform distributions.
(c) The same distribution as in Fig.~\ref{fig:sm_distrib_all_models}(b), but with a basal division rate $k_d^0=k_d$.
(d) Coefficients of variation ($\mathrm{CV}$) as a function of $\gamma$ for different basal division rates $k_d^0$. Note that we have used $\gamma\to\gamma'=\gamma/2$, as in Figs.~\ref{fig:sm_distrib_all_models}(b) and \ref{fig:statistics}(d).
}
\label{fig:suppl_basal}
\end{figure}

If a basal division rate $k_d^0$ is introduced, the results are relatively straightforward---the system simply transitions from the end-growth case to the all-growth case. Within our matrix equation framework---e.g., in the one-end growth model---the total growth rate becomes: $g(n)=k_d^0n + k_d$.
This modification allows all cells to divide at a small but finite rate $k_d^0$.
In the near-deterministic limit $\gamma\gg 1$, the distribution transitions from a uniform distribution in the range $[N_\mathrm{min}, 2 N_\mathrm{min}]$ to the predicted log-uniform distribution in the all-cell growth model.
To check whether some of the key qualitative results in the end-growth model are still preserved, we set $N_\mathrm{min}=1$ and observe that the steady-state distribution remains largely unchanged until $k_d^0$ becomes comparable to $k_d$. Figure~\ref{fig:suppl_basal}(c) shows the distributions $p_n$ corresponding to Fig.~\ref{fig:sm_distrib_all_models}(b), but with a basal division rate $k_d^0=k_d$. While the peaks persist, $p_n$ increasingly resembles the all-growth case [Fig.~\ref{fig:sm_distrib_all_models}(a)] with long exponential tails.
Regarding relative errors, the coefficients of variation remain $<1$ and gradually approach the all-growth results as $k_d^0$ increases.
{We expect the key factor of the relevance of the basal growth rate is whether growth is dominated by basal growth or end-growth, i.e., for the most common cluster sizes $n$, whether $n k_d^0$ or $2 k_d$ is larger. This means that if there is a basal growth rate, as cluster sizes get larger and larger, we would expect that the system looks more and more like the all-growth case. We see this in Fig.~\ref{fig:suppl_basal}(d) for $k_d^0 = 0.1 k_d$: As $\gamma \ll 1$, the $\mathrm{CV}$ transitions from the end-growth value and starts to increase, presumably reaching eventually to $\mathrm{CV}=1$ as $\gamma$ becomes small enough.}

{
\section{Conditional survival probability}\label{app:conditional_survival}
For systems that do not reach a steady state, an ensemble initialized from a single cell is more directly comparable to experiments. Accordingly, the first conditional survival probability $u_1(t)$ is of particular interest.
Solving Eq.~\eqref{eq:KBE} analytically for $u_1$, however, requires a closed-form expression for the total hazard rate $\lambda(n)$. While such an expression is not available in full generality, the problem can be solved explicitly under the simplest assumption $q_{mn}=k_b$, corresponding to our one-dimensional all-cell growth model, using a generating-function approach.
}

{
With $q_{mn}=k_b$, the total hazard rate is $\lambda(n)=(n-1)k_b$, and Eq.~\eqref{eq:KBE} reduces to
\begin{equation}
    \partial_t u_n=-(n-1)k_b u_n +n k_d(u_{n+1}-u_n).
\end{equation}
We write $u_n(t)=a_n(t)\mathrm{e}^{k_b t}$, which yields
\begin{equation}
    \partial_t a_n=-n(k_b+k_d) a_n + nk_da_{n+1}.
\end{equation}
We define the generating function
\begin{equation}
    G(z,t)\coloneq\sum_{n\geqslant1} a_n(t) z^n.
\end{equation}
Multiplying the evolution equation for $a_n$ by $z^n$ and summing over $n\geqslant 1$, we obtain
\begin{equation}
    \partial_t G = -(k_b+k_d)z\partial_z G +k_d\partial_z G -(k_d/z)G.
\end{equation}
Solving this first-order PDE with the initial condition $G(z,0)=\sum_{n\geqslant1} z^n$, which follows from the definition $u_n(0)=1$ and hence $a_n(0)=1$, gives
\begin{equation}
    G(z,t)=\frac{(k_b+k_d)z}{k_b\mathrm{e}^{(k_b+k_d)t}+k_d-(k_b+k_d)z}.
\end{equation}
The coefficients $a_n(t)$ are then obtained by expanding $G(z,t)$ as a power series in $z$:
\begin{equation}
    G(z,t)=\sum_{n\geqslant1}\left(\frac{k_b+k_d}{k_b \mathrm{e}^{(k_b+k_d)t}+k_d}\right)^n z^n.
\end{equation}
Consequently,
\begin{equation}
    u_n(t)=a_n(t)\mathrm{e}^{k_bt} = \left(\frac{1+\gamma}{1+\gamma\mathrm{e}^{(1+\gamma)k_d t}}\right)^n \mathrm{e}^{k_b t},
\end{equation}
which depends explicitly on both $k_b$ and $k_d$, as shown in Fig.~\ref{fig:suppl_conditional_survival}. We can verify that the total survival probability, averaged over the steady-state distribution [Eq.~\eqref{eq:app_pn_all}], is
\begin{equation}
    S(t)=\sum_{n\geqslant1} p_n\cdot u_n(t)=\mathrm{e}^{-k_d t}.
\end{equation}
}

\begin{figure}
\hspace{-0.2cm}
\includegraphics[width=0.48\textwidth]{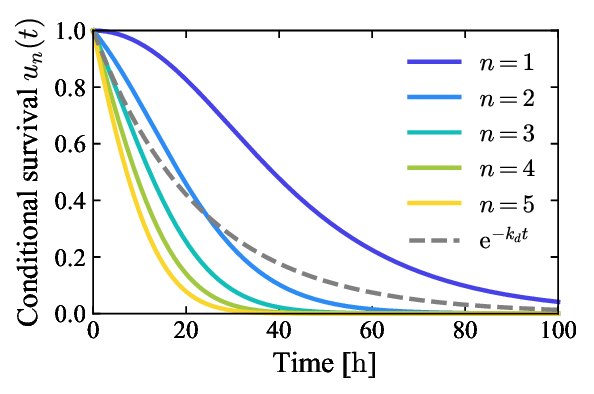}
\caption{{Conditional survival probability $u_n(t)$ for different initial cluster sizes $N(0)=n$ in the one-dimensional all-cell growth model, using the estimated break and division rates for germline cysts: $k_b = 0.02\,\mathrm{h}^{-1}$ and $k_d=(\ln2/16)\,\mathrm{h}^{-1}$. The gray dashed line shows the total survival probability $S(t)=\mathrm{e}^{-k_d t}$, averaged over the corresponding steady-state distribution in Eq.~\eqref{eq:app_pn_all}.}}
\label{fig:suppl_conditional_survival}
\end{figure}

\section{Simulation details}\label{app:mc}
\subsection{Active-particle simulations}\label{app:AOUP_simulation}
We simulate the dynamics of $N$ active Ornstein-Uhlenbeck particles (AOUPs) described by Eq.~(1) using the forward Euler method (Euler-Maruyama method~\cite{kloeden1992stochastic}) as in Ref.~\cite{fodor2016far}.
We integrate the dynamics with a time step $\Delta t=10^{-3}\,\mathrm{h}$ over a total simulation time $T=10^3\,\mathrm{h}$.
{The variances of the stretched lengths of the springs $\langle\delta\ell_n^2\rangle$ are calculated from $4800$ independent simulation runs.
We have verified that $\langle \delta\ell_n^2\rangle$ does not depend on the position $n$ within the chain when $N$ is large in the simulations. Therefore, $\langle\delta\ell_{N/2}^2\rangle$ for the central link is used to represent the variance in Fig.~\ref{fig:mechanics}(b).}

To measure the mean escape time $\tau_\mathrm{esc}$, we first allow the chain to relax to its steady state by running the simulation for $10^3\,\mathrm{h}$. After relaxation, we track the stretched length of the central link $\delta\ell_{N/2}$, starting the timer when it approaches zero, i.e., $|\delta\ell_{N/2}|<\varepsilon=0.1\,\textrm{\textmu}\mathrm{m}$ in our simulations.
The break time $\tau_\mathrm{esc}$ is recorded as the time at which $\delta\ell_{N/2}$ reaches the critical break length $\delta\ell_b$.
The final results shown in Fig.~\ref{fig:mechanics}(c) are averaged over $10^3$ independent simulations.

\begin{table}
\caption{\label{tab:param}%
Table of simulation parameters.}
\begin{ruledtabular}
\begin{tabular}{llcc}
Parameter & \qquad Description & Dimension & Value\footnote{These serve as the default parameter values for the AOUP simulations; any deviations from them are explicitly specified.}\\\hline
$N$ & Number of particles & 1 &  $1000$\\
$\Delta t$ & Simulation time step & $T$ & $10^{-3}\,\mathrm{h}$\\
$1/\mu k$ & Spring relaxation time & $T$ & $0.25\,\mathrm{h}$~\cite{jain2020role}\\
$\tau$ & Persistence time & $T$ & $0.5\,\mathrm{h}$~\cite{selmeczi2005cell}\\
$D$ & Diffusion coefficient & $L^2/T$ & $100\,\textrm{\textmu}\mathrm{m^2/h}$~\cite{selmeczi2005cell}\\
$\delta\ell_b$ & Critical break length & $L$ & $5\,\textrm{\textmu}\mathrm{m}$
\end{tabular}
\end{ruledtabular}
\end{table}

\subsubsection{Details of parameter estimates}\label{app:AOUP_params}

The default parameter values used in our model are listed in Table~\ref{tab:param}.
These values have been calibrated using experimental data as discussed in the main paper. We provide a few additional details here. We estimated $D = 100\,\textrm{\textmu}\mathrm{m^2/h}$ and $\tau \approx 0.5\,\mathrm{h}$ from data on HaCaT cells (a human keratinocyte cell line) from Ref.~\cite{selmeczi2005cell} (Fig.~4\textit{E} of that paper). Selmeczi \textit{et al.} show in their paper that the Ornstein-Uhlenbeck model is not a complete representation of their velocity correlation data~\cite{selmeczi2005cell}, so our application here is only to get a reasonable order-of-magnitude.

{We estimate the spring relaxation time $1/\mu k$ based on the results of Jain \textit{et al.}, who calibrated a self-propelled particle model of collective cell migration using their experimental measurements~\cite{jain2020role}. We want to note how our model maps to their calibration briefly, since it might not be immediately obvious. The model of Ref.~\cite{jain2020role} essentially assumes a force balance between a self-propulsion force of the cell $F_M$, a viscous drag, and a cell-cell interaction force, $0 = -\xi v + F_M + f_\textrm{cell-cell}$. In the absence of cell-cell contact, then the cell velocity is given by $v = F_M/\xi$ ($\xi$ is called $\mu$ in Ref.~\cite{jain2020role}; we have changed this to avoid confusion with our mobility $\mu$, which is $1/\xi$). Based on typical cell velocities in their experiment, it's possible to set $F_M/\xi = 30\,\textrm{\textmu}\mathrm{m/h}$. If one cell is colliding into a stationary cell, the cell-cell interaction force is nonzero, and they assume it to be $f_\textrm{cell-cell} = F_C [1-({d}/{2R})]$, where $d$ is the cell-cell distance. Within our model, the force would be $-k\delta \ell$---so with a typical spacing of $\ell_0 = 2R$ of twice the radius of the cell, we see that our $k$ value corresponds to $F_C/2R$ in their model.
By measuring how much a cell is compressed when another cell is pushing against it, Ref.~\cite{jain2020role} estimated the ratio of cell stiffness $F_C$ to the persistent motile force $F_M$ to be approximately $F_C/F_M\approx 5.5$.
This means our value of $k =  F_C/2R \approx {5.5 F_M}/{2R}$, or in our notation, $k/\xi \approx {5.5 F_M}/{2R\xi}$. Using the speed of the cell for $F_M/\xi$ and the value from Ref.~\cite{jain2020role} of $R=20\,\textrm{\textmu}\mathrm{m}$ for MDCK cells, we find $k/\xi =\mu k\approx 4\,\mathrm{h^{-1}}$, or our timescale ${1}/{\mu k} \approx 0.25\,\mathrm{h} = 15\,\textrm{min}$. We note that other papers have come up with different estimates for this timescale~\cite{schoetz2013glassy}, which may vary significantly from cell type to cell type and depending on the context (e.g., cells being on a substrate vs in a 3D aggregate likely have very different frictions).}

{
\subsubsection{2D AOUP simulations}
For the 2D AOUP simulations in Sec.~\ref{sec:2d}, we integrate the dynamics using a time step $\Delta t= 0.01\,\mathrm{h}$ over a total simulation time of $T=10^3\,\mathrm{h}$.
We also perform a time step convergence check by reducing the time step to $\Delta t=0.005\,\mathrm{h}$, which yields results consistent with Fig.~\ref{fig:2d_simulations} within statistical error.
In each run, we track the lineage of a single cluster by randomly selecting one of the two daughter clusters once the cluster has broken into two separate pieces, and continuing the simulation with that daughter cluster.  
We fix the cell division rate $k_d = (\ln 2/16)\,\mathrm{h^{-1}}$, and set cell radius to $R =20\,\textrm{\textmu}\mathrm{m}$, so the natural length of the spring $\ell_0=2R =40\,\textrm{\textmu}\mathrm{m}$. The critical breaking length is set to $\delta\ell_b=6.5\,\textrm{\textmu}\mathrm{m}$ to match germline cysts.
We vary the typical cell speed $v_\textrm{rms}$ by changing $D$ while keeping the persistence time $\tau$ fixed.
The final results shown in Fig.~\ref{fig:2d_simulations} are from $100$ independent simulations.
}

\subsection{Fragmentation-growth simulations}\label{app:rfKMC}
We implement a rejection-free kinetic Monte Carlo (rfKMC) algorithm to simulate the fragmentation-growth processes.
The initial chain length is set to $n_0\gg N_\text{min}$.
For a chain of $N$ particles, there are $N-1$ connections, each with a break rate. If breaking a connection violates the minimum cluster size $N_\textrm{min}$, the break rate is set to zero.
Each particle has a division rate, which is set to zero for particles not allowed to divide (e.g., in the end-growth mechanism).

\begin{figure}[tbp]
\hspace{-0.5cm}
\includegraphics[width=0.45\textwidth]{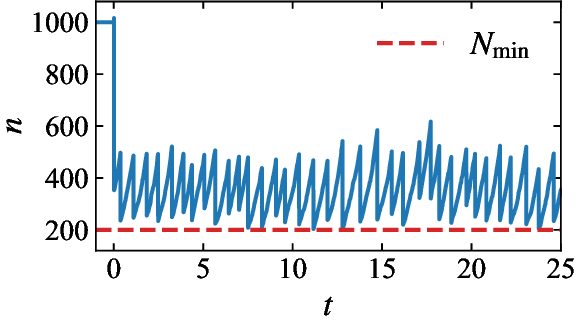}
\caption{A representative time trajectory of chain length $n(t)$ from rfKMC simulation with $N_\textrm{min}=200$ and $\gamma=0.1$. The initial chain length is set to $n_0=1000$, and the simulation begins at $t=0$.
The time unit is $1/k_d$.}
\label{fig:suppl_simulation}
\end{figure}

In our simulations, we fix the division rate $k_d=1$ and vary the ratio $\gamma$ by adjusting the break rate $k_b=\gamma$, i.e., the time unit is $1/k_d$.
In the rfKMC algorithm, the time step for each MC step is calculated as $\Delta t =Q_{k}^{-1}\ln(1/u)$, where the total rate $Q_k$ is the sum of all break and division rates, and $u$ is a uniform random number in $(0,1]$. Each step permits either a fragmentation or a division event. If a particle division occurs, the chain length increases by one. If fragmentation occurs, one of the daughter chains is randomly selected as the new tracked chain. Then, we update all rates for particles and connections according to the specific fragmentation-growth mechanism.
Since the time step $\Delta t$ in rfKMC is not constant, the output is a time series $\{t_0, t_1,t_2,\cdots\}$. To determine the chain size at a specific time $T$, we identify the first time $t_{i+1}>T$, and then record the chain size at $t_i$ as $n(t=T)$.
To obtain the steady-state distribution $p_n$, we must wait for a sufficiently long time $T$ for the chain length $n(t)$ to reach a stable increase-then-drop pattern (as shown in Fig.~\ref{fig:suppl_simulation}), akin to cell size control~\cite{jia2022characterizing, elgamel2024effects}.
The distribution is then generated from the results of $10^4$ independent simulation runs.
{There is an alternative, faster way to obtain the same distribution from simulations. We expect that the steady-state distribution $p_n$ is equivalent to the lineage distribution of the tracked chains as $T\to \infty$, i.e., $\lim_{T\to\infty}T^{-1}\int_0^Tp_n(t)\mathrm{d}t=\lim_{t\to\infty}p_n(t)$~\cite{ocal2024universal}. We can get the lineage distribution by measuring the length $n(t)$ of the tracked chain over a long period and calculating the fraction of time spent at each $n$.
{We have computed these lineage distributions and confirmed that they agree with our distributions shown in Figs.~\ref{fig:statistics}(a), \ref{fig:statistics}(e), and theory.}}

We note that the cluster sizes of germline cysts shown in Fig.~5 of Ref.~\cite{levy2024tug} are smaller than our steady-state predictions using their measured break and division rates, as their system has not yet reached steady state. For instance, the break rate is approximately $0.015\,\mathrm{h^{-1}}$ at E11.5 and E12.5 ovaries (Fig.~4C of Ref.~\cite{levy2024tug}), while the division rate for germline cysts is about $\ln 2/16\,\mathrm{h^{-1}}$. Our steady-state solution predicts a mean cluster size $(1+\gamma)/\gamma\approx 4$, whereas their simulated average cyst sizes are $1.73$ and $3.26$ at E11.5 and E12.5, respectively. If we similarly evolve a single cell only from E10.5 to E11.5 and E12.5, as in Ref.~\cite{levy2024tug}, we obtain mean cluster sizes of approximately $2$ and $3$, which closely match their results. We note the simulations of Ref.~\cite{levy2024tug} also include variability in cell division rates, which we do not include in order to allow us to get a simple analytic answer.

\section{Additional figures}

\bibliography{main}%

\newpage
\onecolumngrid
\vspace{2cm}

\begin{figure*}[h]
\includegraphics[width=0.9\textwidth]{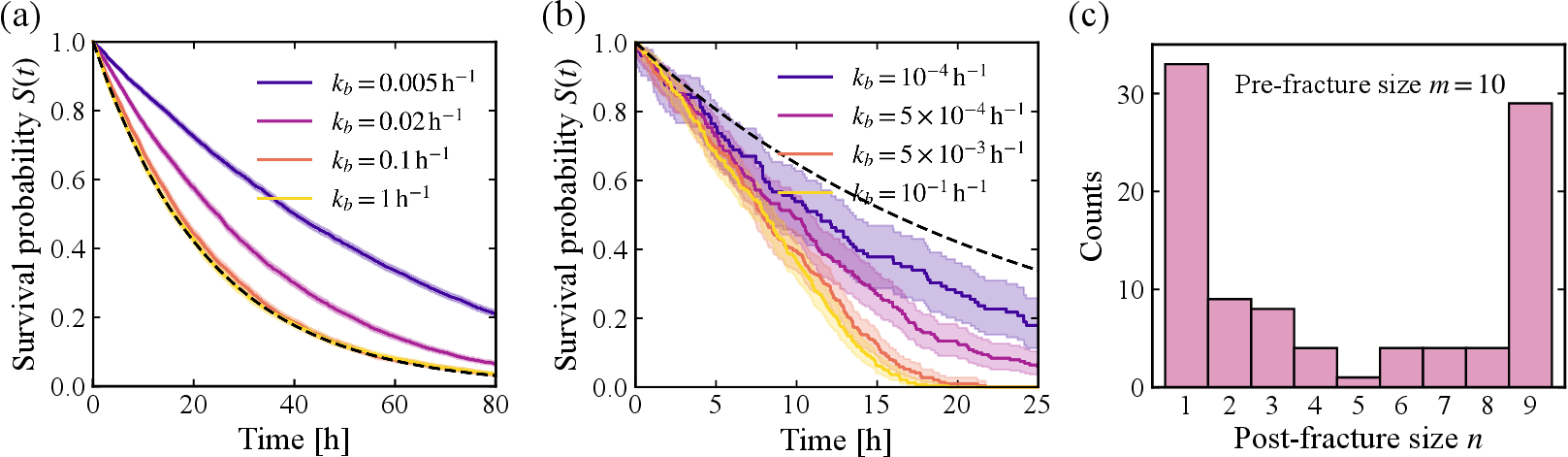}
\caption{
{
(a) Survival probability $S(t)$ from rfKMC simulations of the one-end growth model at fixed division rate $k_d$ with varying break rate $k_b$.
(b) Survival probability $S(t)$ from rfKMC simulations of the all-cell growth model with minimum cluster size $N_\textrm{min}=200$, again varying $k_b$ at fixed $k_d$. Dashed lines are the exponential decay $\mathrm{e}^{-k_d t}$. 
Error bars on $S(t)$ represent 95\% confidence intervals.
(c) Distribution of post-fracture size $n$ for pre-fracture size $m=10$ in 2D AOUP simulations with cell motility $v_\textrm{rms}=22\,\textrm{\textmu}\mathrm{m/h}$.
Division rate $k_d=(\ln 2/16)\,\mathrm{h^{-1}}$.
}}
\label{fig:suppl_survival}
\end{figure*}

\end{document}